\documentclass[aps,pre,showpacs,amsmath,amssymb,longbibliography]{revtex4-1}
\usepackage{amsmath,amssymb,latexsym,epsfig,graphicx,epsf,bm}
\usepackage{graphicx}
\usepackage{float}

\newcommand{\bR}{\bold R}
\newcommand{\br}{\bold r}

\newcommand{\bRa}{{\bf R}_{\rm a}}
\newcommand{\bRb}{{\bf R}_{\rm b}}
\newcommand{\tbR}{{\tilde{\bold R}}}

\newcommand{\be}{\begin{equation}}
	\newcommand{\ee}{\end{equation}}

\newcommand{\Fig}[1]{Figure~\ref{#1}}

\newcommand{\Sex}{{S}_{\rm ex}}

\begin{document}
	\title{Isomorph invariance in the liquid and plastic-crystal phases of asymmetric-dumbbell models}
	\date{\today}
	\author{Eman Attia}\email{attia@ruc.dk}
	\affiliation{\textit{Glass and Time}, IMFUFA, Department of Science and Environment, Roskilde University, P.O. Box 260, DK-4000 Roskilde, Denmark}
	\author{Jeppe C. Dyre}\email{dyre@ruc.dk}
	\affiliation{\textit{Glass and Time}, IMFUFA, Department of Science and Environment, Roskilde University, P.O. Box 260, DK-4000 Roskilde, Denmark}	
	\author{Ulf R. Pedersen}\email{urp@ruc.dk}
	\affiliation{\textit{Glass and Time}, IMFUFA, Department of Science and Environment, Roskilde University, P.O. Box 260, DK-4000 Roskilde, Denmark}

\begin{abstract}
We present a numerical study of the asymmetric dumbbell model consisting of ``molecules'' constructed as two different-sized Lennard-Jones spheres connected by a rigid bond. In terms of the largest (A) particle radius, we report data for the structure and dynamics of the liquid phase for the bond lengths 0.05, 0.1, 0.2, and 0.5, and analogous data for the plastic-crystal phase for the bond lengths 0.05, 0.1, 0.2, and 0.3. Structure is probed by means of the AA, AB, and BB radial distribution functions. Dynamics is probed via the A and B particle mean-square displacement as functions of time and via the rotational time-autocorrelation function. Consistent with the systems' strong virial potential-energy correlations, the structure and dynamics are found to be isomorph invariant to a good approximation in reduced units, while they generally vary considerably along isotherms of the same (20\%) density variation. Even the rotational time-autocorrelation function, which due to the constant bond length is not predicted to be isomorph invariant, varies more along isotherms than along isomorphs. Our findings provide the first validation of isomorph-theory predictions for plastic crystals for which isomorph invariance, in fact, is found to apply better than in the liquid phase of asymmetric-dumbbell models.
\end{abstract}
\maketitle

\section{Introduction}\label{sec1}

This paper presents a systematic study of the asymmetric dumbbell model (ASD) consisting of two different-sized Lennard-Jones (LJ) spheres connected by a rigid bond. The ASD model is designed to be a simple molecular model~\cite{swe09,ing12b,lee16}. Because of the asymmetry, the liquid phase of the model is easily supercooled, i.e., the model does not readily crystallize~\cite{ing12b}. This makes it suitable for numerical studies as a simple, single-component glass-forming model of non-spherically symmetric constituents~\cite{kam97,cho10b,ing12b,fra13a,dau14,fra19}. 

The focus here is not on supercooling and glass formation, however, but on investigating what happens when the bond length is varied in the less viscous liquid as well as plastic-crystal phases. It was previously shown that for the bond length 0.58 (in units of the largest (A) particle's radius), the ASD model obeys the hidden-scale invariance symmetry, which manifests itself in strong virial potential-energy correlations in the thermal-equilibrium constant-volume fluctuations~\cite{ing12b}. This implies the existence of so-called isomorphs, which are lines in the thermodynamic phase diagram along which structure and dynamics in reduced units are invariant to a good approximation~\cite{ing12b,IV,dyr14}. We confirm this below and demonstrate that the existence of isomorphs is robust to bond-length changes. This is done by simulating the ASD model in the liquid phase with bond lengths 0.05, 0.1, 0.2, and 0.5. We also investigate the model's plastic-crystal phase with a focus on the existence of isomorphs; here the largest bond length simulated is 0.3 (for larger bond lengths the system was liquid at the chosen reference state point).

Our investigation reports data illuminating the degree of isomorph invariance of structure and dynamics. This is done by comparing results along isomorphs with results along isotherms of the same density variation. Specifically, the structure is investigated by monitoring the radial distribution function (RDF), while the dynamics is investigated via the mean-square displacement (MSD) as a function of time, as well as the rotational time-autocorrelation function (RAC). We find good isomorph invariance of both structure and dynamics.

The paper presents the first investigation of the consequences of hidden scale invariance for a system of molecules forming plastic crystals. In the plastic-crystal phase the molecules' centers are ordered on a crystalline lattice---in the present case a face-centered cubic lattice---while the molecular orientations vary more or less randomly because the molecules are free to rotate~\cite{bra02,pri10,vis17, azn20,das20}. In this phase we find excellent isomorph invariance, in fact better than in the (isotropic) liquid phase.

\section{Model and Simulation Details}\label{sec2}

The asymmetrical dumbbell (ASD) is a constrained molecular model. It consists of two different-sized Lennard Jones (LJ) particles, a sphere (A) and a slightly smaller one with lower interaction energies (B). If the two spheres are connected by a rigid bond of length = $0.584$, the model mimics toluene~\cite{ped08a,ing12b}. The values of the parameters used below are as follows. For particle A the distance and energy parameters are $\sigma_{AA}=\varepsilon_{AA}=1$ and the mass is $m_{A}=1$; for particle B one has $\sigma_{BB}=0.788$, $\varepsilon_{BB}=0.117$, and $m_{B}=0.195$; for the \emph{AB} interaction one has $\sigma_{AB}=0.894$ and $\varepsilon_{AB}=0.342$~\cite{ped08a}.  

We define the (number) density $\rho$ as the total number of A and B particles ($N=8000$) divided by the simulation box volume $V$; $T$ is below the temperature. The  system is studied by molecular-dynamics simulations in the canonical $NVT$ ensemble using the Nosé-Hoover thermostat~\cite{tildesley}. The simulated system consists of 4000 ASD molecules in a cubic box with periodic boundaries. Simulations are performed using the open-source Roskilde University molecular dynamics software (RUMD) that runs on GPUs (graphics processing units)~\cite{RUMD}. The constant bond-length constraint was implemented using the method of Ref.~\cite{tox09a}. 

The leapfrog algorithm is used with time step  $dt=0.001$ (in A particle LJ units). First, $NVE$ simulations were carried out for $10^{6}$ time steps for equilibration and $10^{6}$ time steps for production runs with time step sizes 0.0005, 0.005, 0.001, 0.002, and 0.01 in order to investigate how constant the energy is at the reference state point defined by $\rho=1.5$ and $T=1.5$. The bond lengths chosen for these $NVE$ simulations were $0.05$ and $0.5$. We concluded that the time step $0.001$ in both cases shows a good energy conservation. For the subsequent $NVT$ simulations generating the data, at each state point the simulations again ran for $10^{6}$ time steps for equilibration and  $10^{6}$ time steps for the production runs. 

For the post analysis of structure and dynamics, the units used are the isomorph-theory's reduced units defined as follows: energy unit: $k_BT$, length unit: $\rho^{-1/3}$, time unit: $\rho^{-1/3}\sqrt{m_A/k_BT}$. Note that these units vary with state point. Exceptions to the use of reduced units apply for density and temperature, which by definition are both unity in reduced units; these quantities, as well as bond lengths, are reported in A particle LJ units.

\section{Essential Isomorph Theory}

Hidden scale invariance is a symmetry of the potential energy function $U(\bR)$ in which the configuration $\bR$ in terms of the particle coordinates is defined by $\bR\equiv (\br_1,\dotso,\br_N)$. If $\bRa$ and $\bRb$ are two same-density configurations, hidden scale invariance is defined by the following mathematical implication (in which $\lambda$ is a uniform-scaling parameter)

\be\label{hsi}
U(\bRa)<U(\bRb)\,\,\Rightarrow\,\, U(\lambda\bRa)<U(\lambda\bRb)\,.
\ee
This symmetry is only rigorously obeyed if $U(\bR)$ is an Euler-homogeneous function plus a constant, but it applies to a quite good approximation for many other systems, e.g., single- and multicomponent LJ systems, the Yukawa and bi-Yukawa pair potentials, the EXP and Morse pair potentials, Mie potentials involving various exponents, etc.~\cite{sch14,IV,ing12b,dyr14,EXPII,cas19a}. Equation~(\ref{hsi}) may also apply for molecular models like the ASD model. The derivations given below apply in the case of constant-length rigid bonds, i.e., \textit{not} scaled with $\lambda$. Equation~(\ref{hsi}) then refers to a uniform scaling of the center-of-mass coordinates with molecular orientations and sizes kept unchanged. 

The excess entropy $\Sex$ is defined as the entropy minus that of an ideal gas at the same density and temperature. Any state point of the thermodynamic phase diagram is fully characterized by the two thermodynamic variables $\rho$ and $T$, but other variables may of course also be used to characterize a state point, for instance density and average potential energy. We define the \textit{microscopic excess-entropy} function $\Sex(\bR)$ by~\cite{sch14}\vspace{-3pt}
\be\label{sex_def}
\Sex(\bR)
\,\equiv\, \Sex(\rho,U(\bR))\,.
\ee
Here the function $\Sex(\rho,U)$ is the excess entropy of the state point $(\rho,U)$ in which $U$ is the average potential energy. It follows from statistical mechanics that $\Sex(\bR)$ is proportional to the logarithm of the number of configurations with the same density and potential energy as $\bR$~\cite{lan58}. Inverting Equation~(\ref{sex_def}) leads to

\be\label{any}
U(\bR)
\,=\,U(\rho,\Sex(\bR))\,,
\ee
in which $U(\rho,\Sex)$ is the average potential energy of the state point of density $\rho$ and excess entropy $\Sex$. 

It can be shown that Equation~(\ref{hsi}) implies the function $\Sex(\bR)$ is scale invariant, i.e., $\Sex(\lambda\bR)=\Sex(\bR)$~\cite{sch14}. In this case, $\Sex(\bR)$ depends only on the configuration's so-called reduced coordinate vector $\tbR\equiv\rho^{1/3}\bR$ (which is of course invariant upon a uniform scaling):

\be\label{Sex_rc}
\Sex(\bR)
\,=\,\Sex(\tbR)\,.
\ee
Consequently, Equation~(\ref{any}) becomes

\be\label{fundeq}
U(\bR)
\,=\,U(\rho,\Sex(\tbR))\,.
\ee
All identities of the isomorph theory may be derived from Equation~(\ref{fundeq})~\cite{sch14}. In particular, Equation~(\ref{fundeq}) implies strong correlations between the constant-volume fluctuations of the virial $W$ and the potential energy $U$. Thus, the equilibrium deviations from the average of these two variables, $\Delta W$ and $\Delta U$, obey~\cite{ped08,I,II}

\be\label{Delta_eq}
\Delta W\,\cong\,\gamma\,\Delta U\,.
\ee
Here the so-called density-scaling exponent $\gamma$ is defined and characterized by~\cite{IV} 

\be\label{gamma}
\gamma
\,\equiv\, \left(\frac{\partial\ln T}{\partial\ln \rho}\right)_{\Sex}
\,=\,\frac{\langle\Delta U \Delta W \rangle}{\langle(\Delta U)^2\rangle}\,.
\ee
The second (generally valid) identity allows one to calculate $\gamma$ from the constant-volume thermal-equilibrium canonical-ensemble fluctuations at the state point in question. 

Using the identity $T=(\partial U /\partial\Sex)_\rho$, a first-order Taylor expansion of Equation~(\ref{fundeq}) leads~\cite{sch14} to
\be\label{firstord}
U(\bR)\,\cong\, U(\rho,\Sex)\, +\, T(\rho,\Sex) \left(\Sex(\tbR)-\Sex\right)\,.
\ee
Consider two state points $(\rho_1,T_1)$ and $(\rho_2,T_2)$ with average potential energies $U_1$ and $U_2$ and the same excess entropy $\Sex$, and suppose that $\bR_1$ and $\bR_2$ are equilibrium configurations of the two state points with the same reduced coordinates, i.e.,  $\rho_1^{1/3}\bR_1=\rho_2^{1/3}\bR_2\equiv \tbR$. In that case, by elimination of $\Sex(\tbR)-\Sex$, Equation~(\ref{firstord}) implies that with $T_1\equiv T(\rho_1,\Sex)$ and $T_2\equiv T(\rho_2,\Sex)$ one has

\be\label{isom}
\frac{U(\bR_1)-U_1}{k_B T_1}\,\cong\,\frac{U(\bR_2)-U_2}{k_B T_2}\,.
\ee
This implies that 

\be\label{isomeq}
e^{-{U(\bR_1)}/{k_B T_1}}\,\cong\, C_{12}\,e^{-{U(\bR_2)}/{k_B T_2}}\,
\ee
in which $C_{12}$ is a constant. Equation~(\ref{isomeq}) is the original 2009 definition of isomorphic state points~\cite{IV}, which stated that the canonical probabilities of configurations that scale uniformly into one another are identical along an isomorph (the value of the constant $C_{12}$ is irrelevant because probabilities are normalized). This identity of the probabilities of scaled configurations implies that $\Sex$ is constant along an isomorph. In fact, the 2014 version of the isomorph theory that introduced Equation~(\ref{hsi}) \textit{defined} isomorphs as lines of constant excess entropy in the thermodynamic phase diagram and showed from this that the excess entropy is an isomorph invariant~\cite{sch14}. 

It can be shown, either from Equation~(\ref{isomeq})~\cite{IV} or from Equation~(\ref{fundeq})~\cite{sch14}, that the dynamics at two isomorphic state points are identical in the ``same movie'' sense: Filming the molecules' motion at one state point results in the same movie at a different, isomorphic state point---except for a uniform scaling of space and time. This implies several dynamic isomorph invariants and that the reduced-unit structure is isomorph invariant in reduced units. 

How does one decide whether a given system is expected to have good isomorphs? According to Equation~(\ref{Delta_eq}) this is the case when the virial and potential-energy fluctuations are highly correlated, which can be investigated by evaluating their Pearson correlation coefficient,
\be\label{R}
R\,=\,\frac{\langle\Delta U \Delta W \rangle}{\sqrt{\langle(\Delta U)^2\rangle\langle(\Delta W)^2\rangle}}\,.
\ee
A system is ``R-simple'', i.e., simple in the sense of having good isomorphs and thus an essentially one-dimensional thermodynamic phase diagram, if $R>0.9$ at the state points of interest~\cite{ped08,I,sch14}. This is a somewhat arbitrary criterion, though, in the sense that the property of  good isomorph invariance depends on the quantity in question---in practice some reduced-unit quantities are more isomorph invariant than others. Note also that how invariant a given property is, of course, depends on how large a density range that is explored.

\section{Results for the Liquid Phase}

\begin{figure}[H]
    \centering	
    \includegraphics[width=5.4cm]{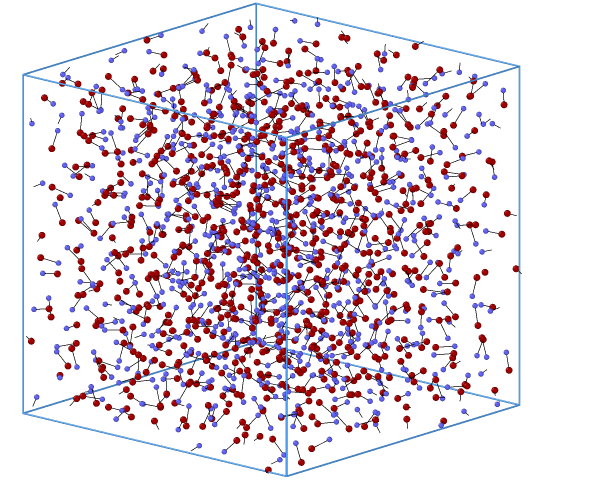}
    \includegraphics[width=5.2cm]{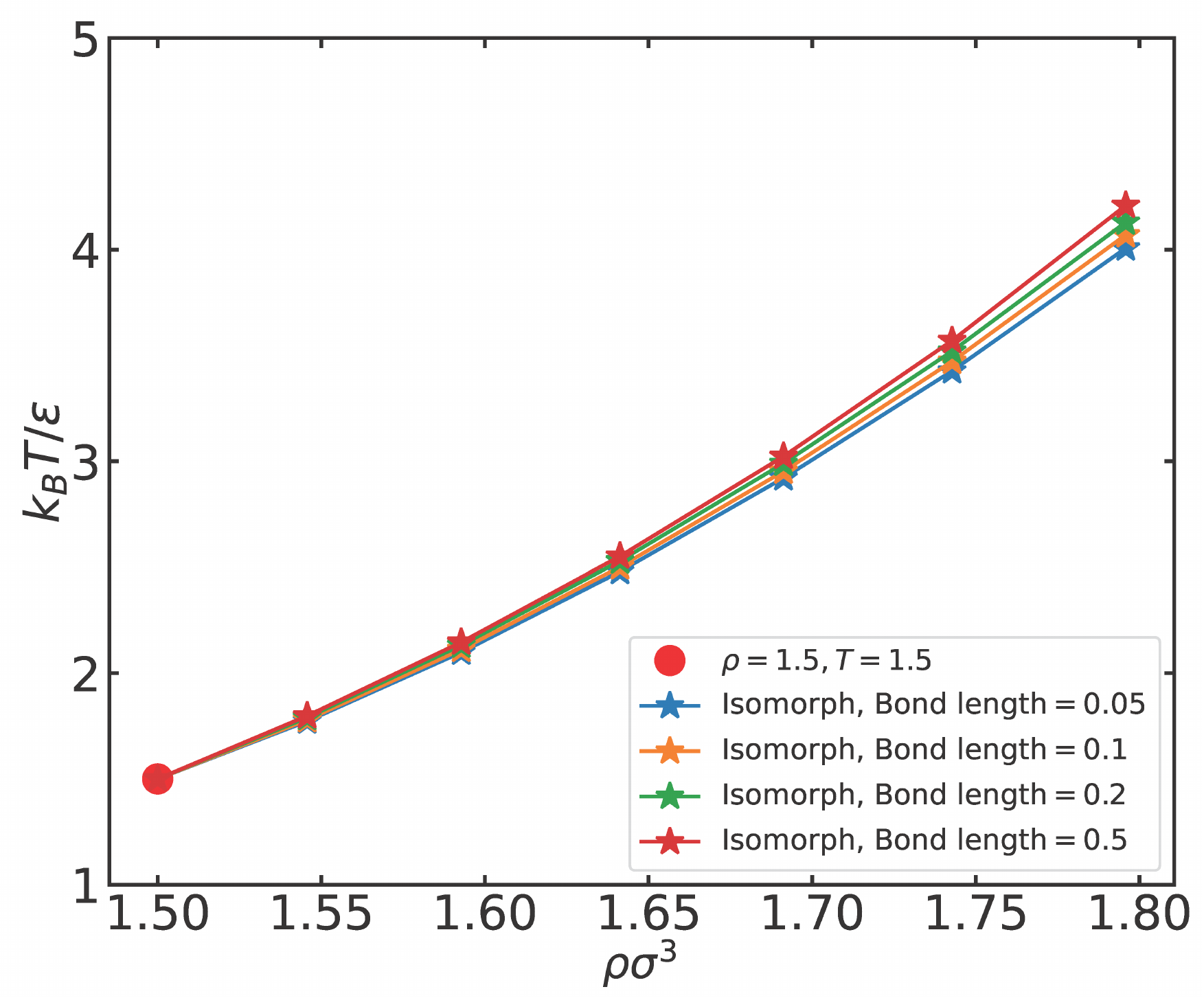}
    \includegraphics[width=5.2cm]{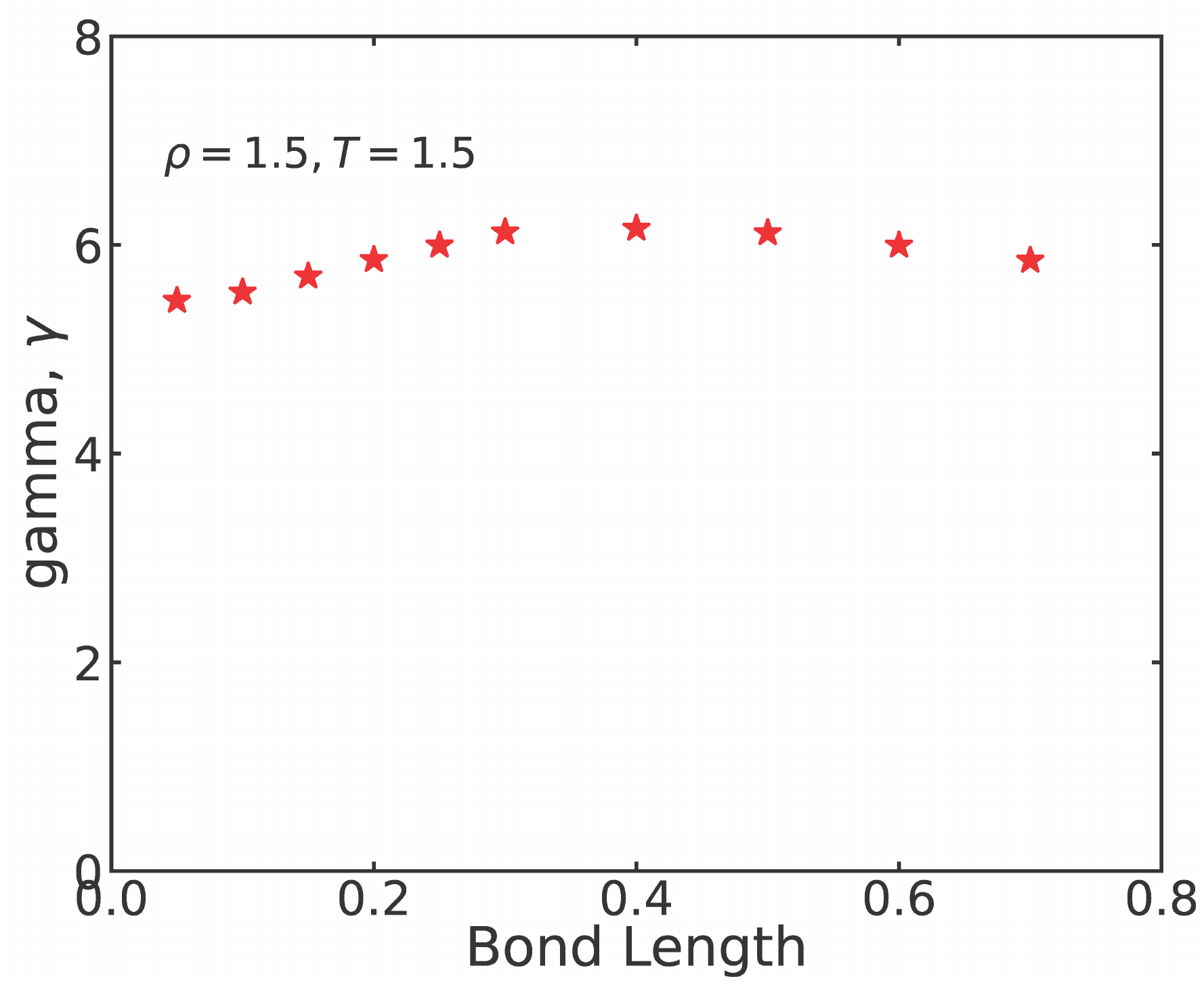}
  	\caption{(\textbf{a}) Snapshot of a liquid configuration at $\rho=0.8, T=0.8$. The bond length between particle A (red) and B (blue) is in this case $0.58$ (in A particle units) ~\cite{ing12b}.
	(\textbf{b}) Four isomorphs traced out in the liquid regime of the ASD thermodynamic phase diagram using the fourth-order Runge--Kutta (RK4) method with density step size $0.01$. Each isomorph is traced out with a different bond length (from $0.05$ to $0.5$), starting from the reference state point marked in red. The isomorphs are qualitatively similar, but we observe a slight increase in the temperatures of the traced state points with increasing bond length.  
  	(\textbf{c}) The variation of the density-scaling exponent $\gamma$ (Equation~(\ref{gamma})) at the reference state point $\rho=1.5, T=1.5$ of each isomorph for the different bond lengths. Data for a few extra bond lengths have been added in order to have a better view of the $\gamma$ variation. We see that $\gamma$ has a maximum around bond length 0.3--0.4, indicating a change of physics here.}
  	\label{fig1} 
\end{figure}

In this section, we investigate the liquid phase for the bond lengths 0.05, 0.1, 0.2, and 0.5. First, however, Figure~\ref{fig1}a shows a snapshot of the ASD liquid at the bond length previously studied by Ingebrigtsen {et al.}~\cite{ing12b} ($0.58$)~\cite{ing12b} at a liquid state point. The A particles are red, the B particles are blue. We see the typical disorder of a liquid configuration. Reducing the bond length results in a system that is increasingly like the atomic LJ liquid, but even for the smallest bond length ($0.05$) we still find typical molecular-liquid behavior (see below).

Isomorphs, i.e., curves of constant excess entropy $\Sex$, are traced out by numerically integrating Equation~(\ref{gamma}). In this way one avoids thermodynamic integration to determine $\Sex$ throughout the phase diagram. Specifically, the last term of Equation~(\ref{gamma}) specifies how the density-scaling exponent $\gamma$ may be calculated from an $NVT$ simulation at the state point in question. Integrating the first-order differential equation defined by the second equality sign in Equation~(\ref{gamma}) in order to determine how temperature varies with density along an isomorph, is in principle straightforward. The highly accurate fourth-order Runge--Kutta (RK4) integration method was recently implemented and used for this~\cite{att21,cas21}. We used that method here with a density change of 1\%. For all four bond lengths we started the integration at the reference state point $(\rho,T)=(1.5,1.5)$. Thus, all four isomorphs have this state point in common.  Table \ref{tab1} gives the pressure, density-scaling exponent, and virial potential-energy correlation coefficient for the four bond lengths at the reference state point. 

\begin{table}[H]
	\caption{\label{tab1} Thermodynamic parameters at the liquid reference state point, giving density, temperature, and pressure (in A particle LJ units), as well as the density-scaling exponent $\gamma$ and the virial potential-energy correlation coefficient $R$.}
		\setlength{\tabcolsep}{3.6mm}\begin{tabular}{cccccc}
			\toprule
		\textbf{Bond Length [}$\boldsymbol{\sigma}_{\textbf{\emph{AA}}}$\textbf{]} & $\boldsymbol{\rho}$ \textbf{[}$\textbf{1/}\boldsymbol{\sigma}_{\textbf{\emph{AA}}}^\textbf{3}$\textbf{]} & $\textbf{\emph{T}}$ \textbf{[}$\boldsymbol{\varepsilon}_{\textbf{\emph{AA}}}\textbf{/$k_B$}$\textbf{]} & $\textbf{\emph{p}}$ \textbf{[}$\boldsymbol{\sigma}_{\textbf{\emph{AA}}}^{\textbf{-3}}\boldsymbol{\varepsilon}_{\textbf{\emph{AA}}}$\textbf{]} & $\boldsymbol{\gamma}$ & $\textbf{\emph{R}}$ \\
			0.05 & 1.5 & 1.5 & 1.09 & 5.47  &  0.85 \\
			0.10 & 1.5 & 1.5 & 1.05 & 5.55 &  0.85  \\
			0.20 & 1.5 & 1.5 & 1.61 & 5.94 &  0.89 \\
			0.50 & 1.5 & 1.5 & 7.50 & 6.12 &  0.96 \\
		\end{tabular}
\end{table}

The isomorphs are not very different (Figure~\ref{fig1}b), but we note that larger bond lengths result in slightly larger temperatures at a given density. This reflects larger density-scaling exponents $\gamma$ for the larger bond lengths, as is evident from Figure~\ref{fig1}c which shows $\gamma$ at the reference state point for several bond lengths. A value of $\gamma$ between 5 and 6 is typical for atomic LJ systems~\cite{I,II}. In fact, in the liquid phase a density-scaling exponent between 5 and 6 is not only typical of atomic LJ system, but also for diatomic LJ systems that often have values somewhat higher than the atomic case~\cite{goe08a,gal11}. Interestingly,  density-scaling exponent relatively close to 6 are often found in experiments~\cite{rol05,gun11,adr16}. We note a slight drop of $\gamma$ at the largest bond length studied, but the overall $\gamma$ variation is just 15\% while, as we shall see, the physics varies considerably when the bond length is changed from 0.05 to 0.5. 

\begin{figure}[H]\centering
	\includegraphics[width=6.5cm]{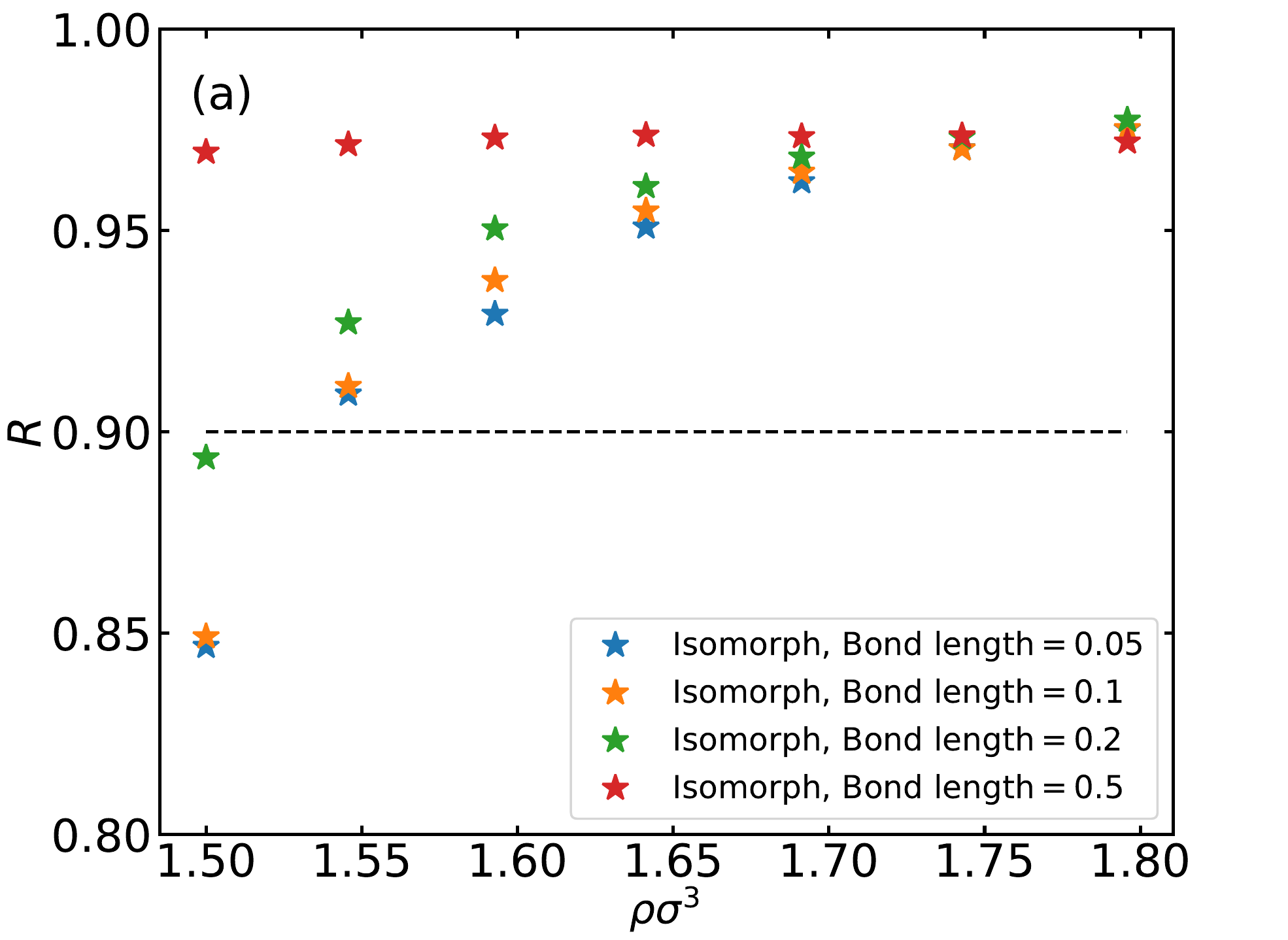}
	\includegraphics[width=6.5cm]{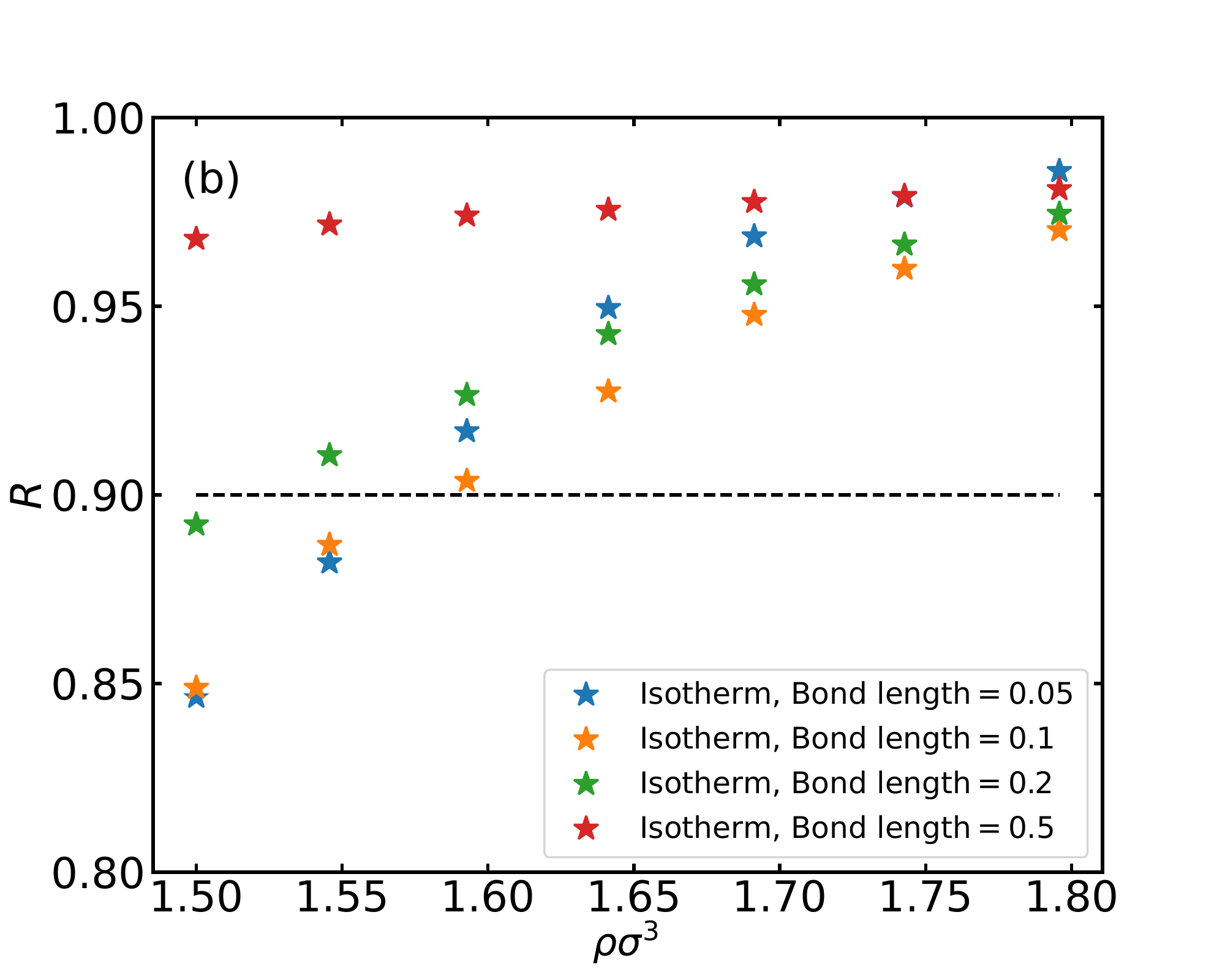}
    \caption{(\textbf{a}) Variation of the virial potential-energy correlation coefficient $R$ along each isomorph plotted for the four bond lengths. 
    (\textbf{b}) The variation of $R$ plotted as a function of density along along the $T=1.5$ isotherm for the four bond lengths.}
	\label{fig2}  
\end{figure}

As mentioned, the isomorph theory only applies for systems with strong virial potential-energy correlations. To ensure that this requirement is obeyed, we evaluated the correlation coefficient $R$ (Equation~(\ref{R})) along the isomorphs and along $T=1.5$ isotherms. The results are shown in Figure~\ref{fig2} as functions of the density in which the traditional $R=0.9$ pragmatic limit for isomorph theory to apply is marked by horizontal dashed lines. We see that the majority of state points are above this line, suggesting that the isomorph theory applies. The strongest correlations are found for the largest bond length studied (0.5).

\begin{figure}[H]
    \centering	\includegraphics[width=8.6cm]{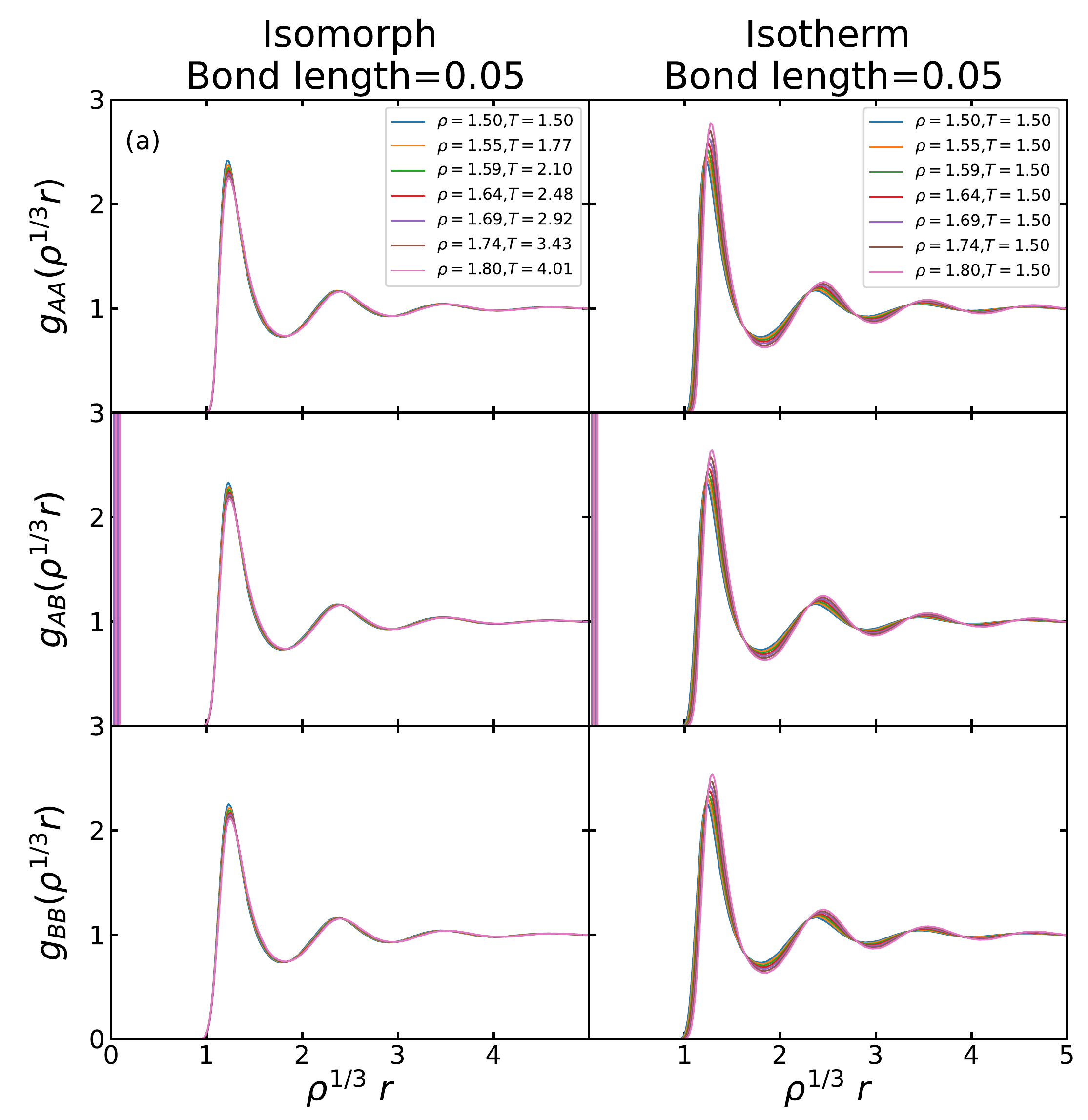}
	\includegraphics[width=8.6cm]{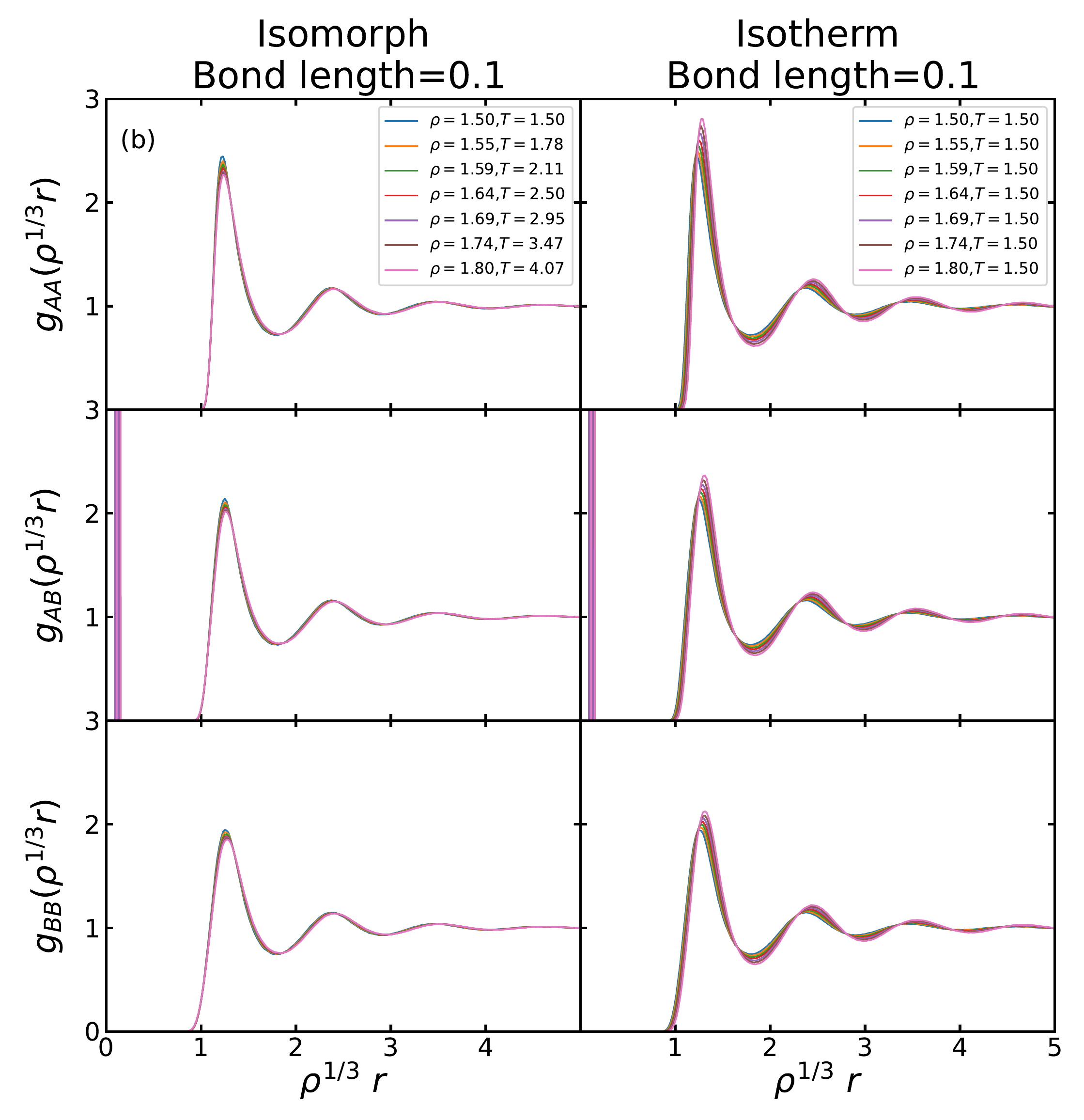}
	\includegraphics[width=8.6cm]{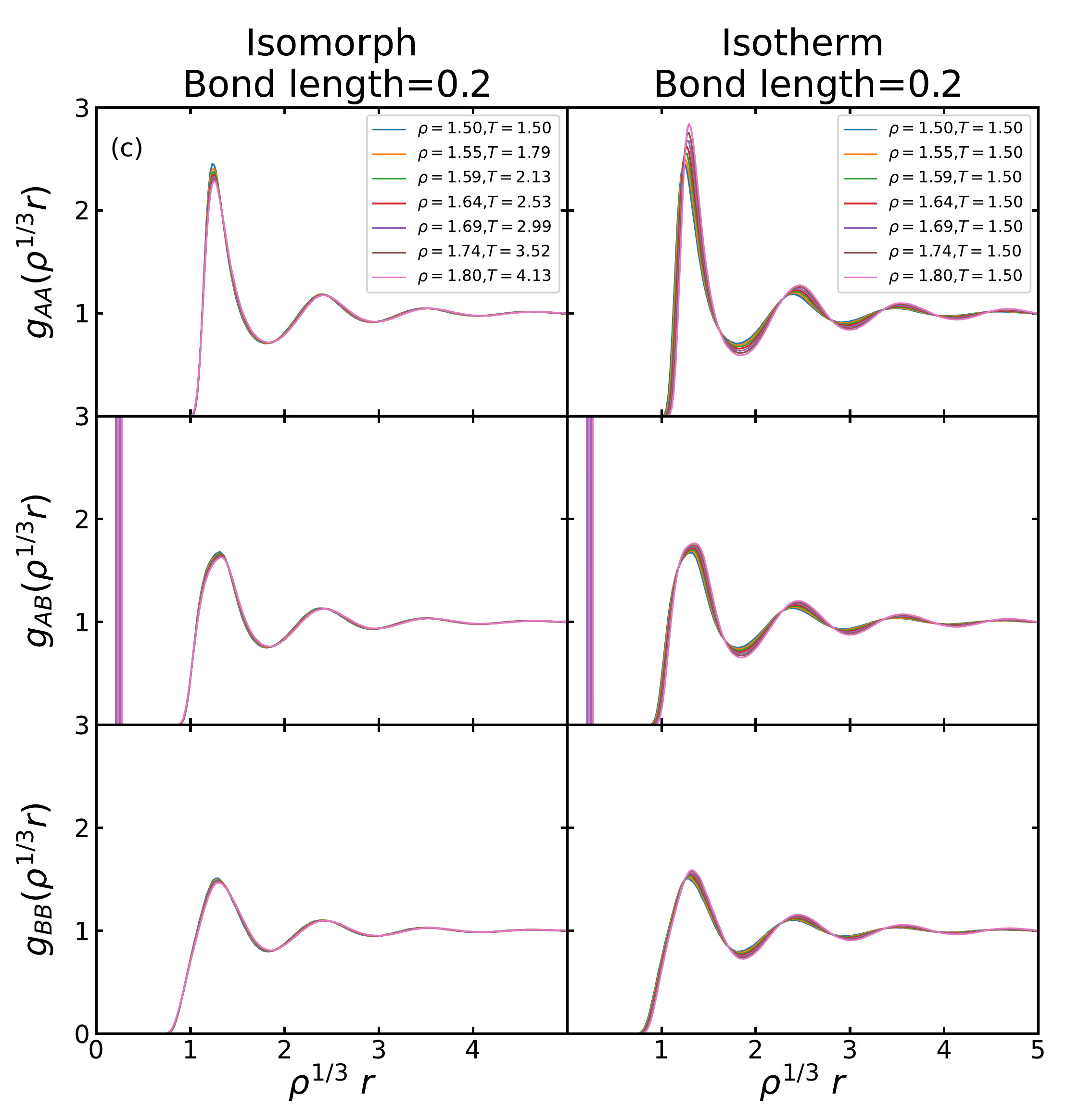}
	\includegraphics[width=8.6cm]{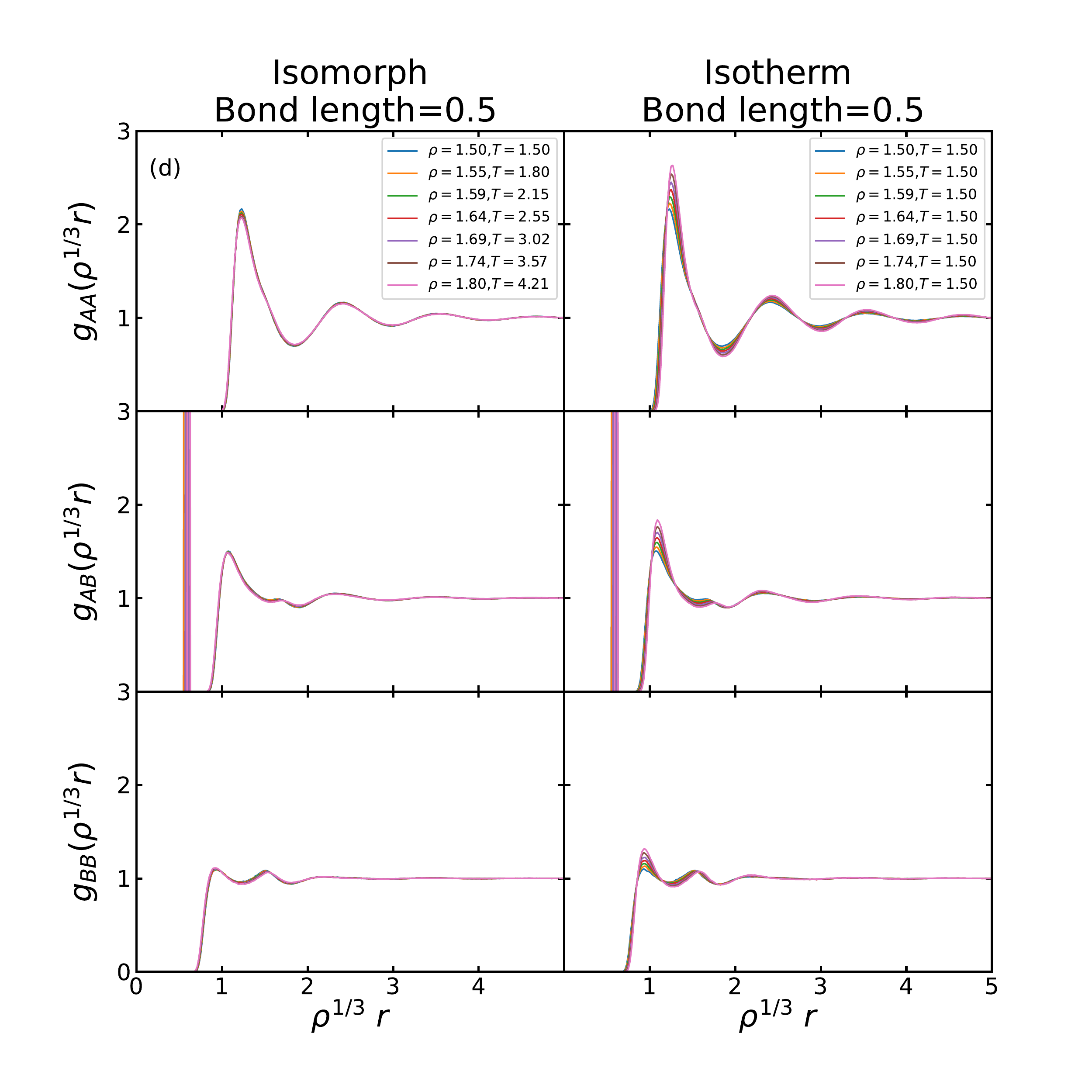} 
	\caption{AA, AB and BB radial distribution functions (RDF) at the reduced pair distance $\rho^{1/3}r$ where $r$ is the pair distance. (\textbf{a}) The RDFs along the isomorph for the bond length 0.05; for comparison the same RDFs are shown along the reference-state-point isotherm of the same (20\%) density variation. We see good, but not perfect invariance along the isomorph, unlike along the isotherm. The thick vertical line in the AB RDF comes from the fixed bond length, which in reduced units varies with density. 
	(\textbf{b}--\textbf{d}) show similar plots for bond lengths 0.1, 0.2, and 0.5, respectively. There is good isomorph invariance of all three RDFs in comparison to their isotherm variation. We note that the first peak of the BB RDF gets lower as the bond length increases. This reflects an increased spread of the B particle positions relative to each other. At the same time, the AB RDFs are lowered and their second peak almost disappears. }
	\label{fig3}  
\end{figure}

To investigate to which degree there is isomorph invariance of the reduced-unit structure we performed simulations of the three radial distribution functions (RDF) $g(r)$ defined by the AA, AB, and BB particle pairs~\cite{tildesley} along the isomorphs. The results are shown in Figure~\ref{fig3}, which for comparison also shows data for simulations carried out at the reference-state point isotherm with the same (20\%) density variation, i.e., for densities between that of the reference state point, 1.5, and 1.8. All three RDFs are nicely invariant along the isomorphs compared to their variation along the isotherms. We note a slight variance at the first peak maximum. This deviation from isomorph theory has often been observed; it is due to the fact that in the direction of increasing $\gamma$ along an isomorph, the system becomes more hard-sphere like. Consequently, the left hand side of the first peak diminishes when moving along an isomorph in the direction of increasing $\gamma$. To ``compensate'' for this effect and maintain a constant overall coordination number (defined as the RDF integral over the first peak), the peak maximum increases. The variance of the first peak maximum has recently been put into a consistent framework by proofs that the so-called bridge function is isomorph invariant~\cite{cas21}.

Focusing henceforth on the isomorph RDFs, we note that the AA RDF does not change very much as the bond length is increased. In fact, this RDF looks much like that of the standard single-component LJ liquid, which is also very similar to that of, e.g., the hard-sphere, Yukawa, and  inverse-power-law liquids~\cite{dyr16}. The observed variation of the AB and BB RDFs can be interpreted as a consequence of the near invariance of the AA particle RDF as the bond length is increased: The structure of the ASD liquid may be thought of as primarily determined by the A particles, which behave much like standard LJ particles that are not really bothered by the B particles. This is because the B particles are smaller and have significantly lower interaction energy parameters. In this picture, the B particles are to a significant degree ``slaves'' of the A particles, and as a consequence one expects what is observed in Figure~\ref{fig3}: The BB RDF changes significantly with increasing bond length, and at bond length 0.5 it is almost constant. This is because the B particles place themselves in many possible positions around the A particles that, as mentioned, more or less have the RDF of a single-component LJ liquid. This effect is present at all bond lengths, but for the smaller bond lengths it does not give rise to any significant spread of the BB RDF because a given B particle is constrained to be close to the A particle of the same molecule. Indeed, the BB particle RDF is very similar to the AA RDF for the small bond length 0.05. This ``B slaving A'' picture is confirmed by the AB RDF, which (except for the vertical line coming from the intramolecular AB bond) gives data for the AB correlations between different ASD molecules. The AB RDF is also significantly smeared out as the bond length increases, but less so than the BB RDF because the relative order of the A particles is partly inherited by the AB RDF.

\begin{figure}[H]
    \centering	
    \includegraphics[width=7.5cm]{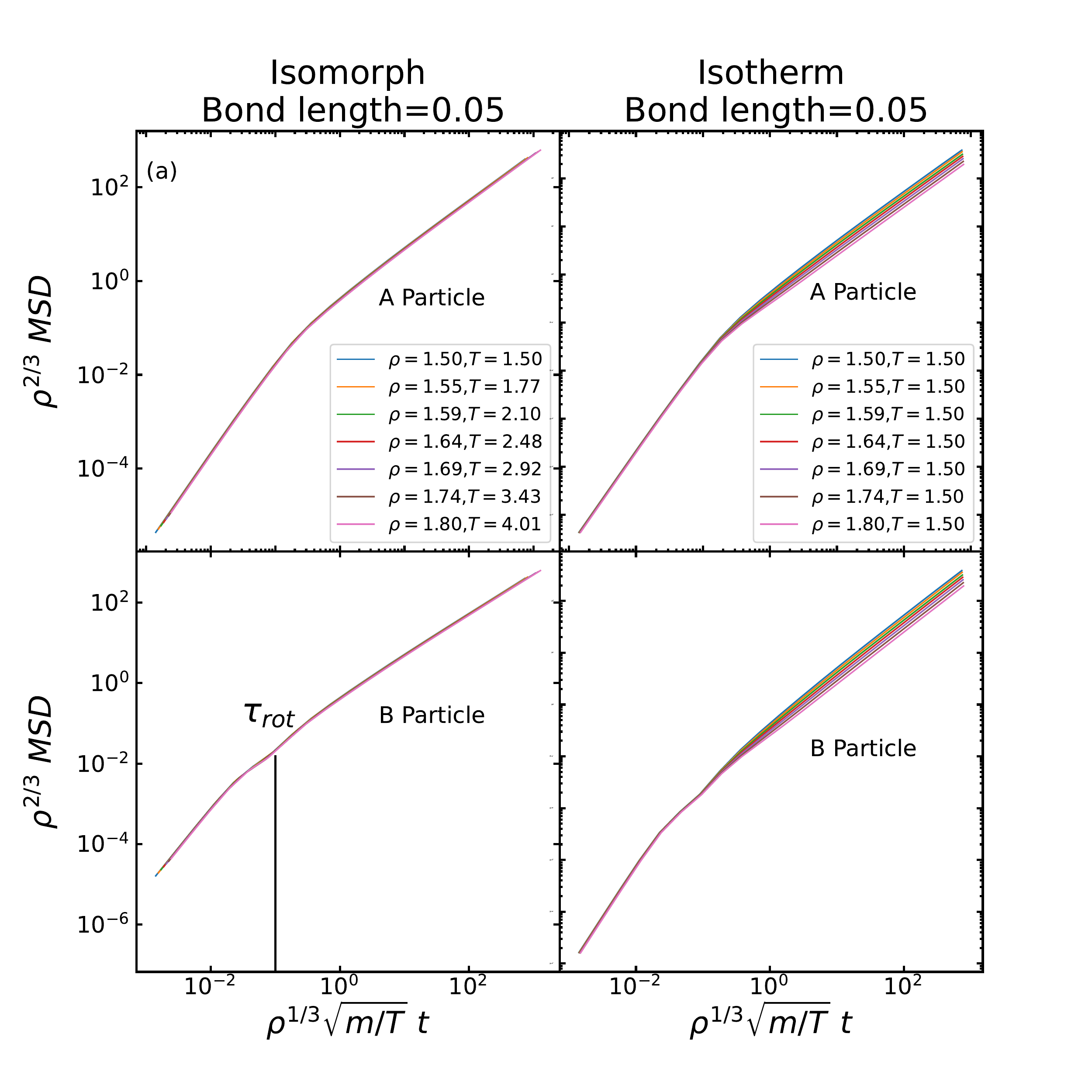}
	\includegraphics[width=7.5cm]{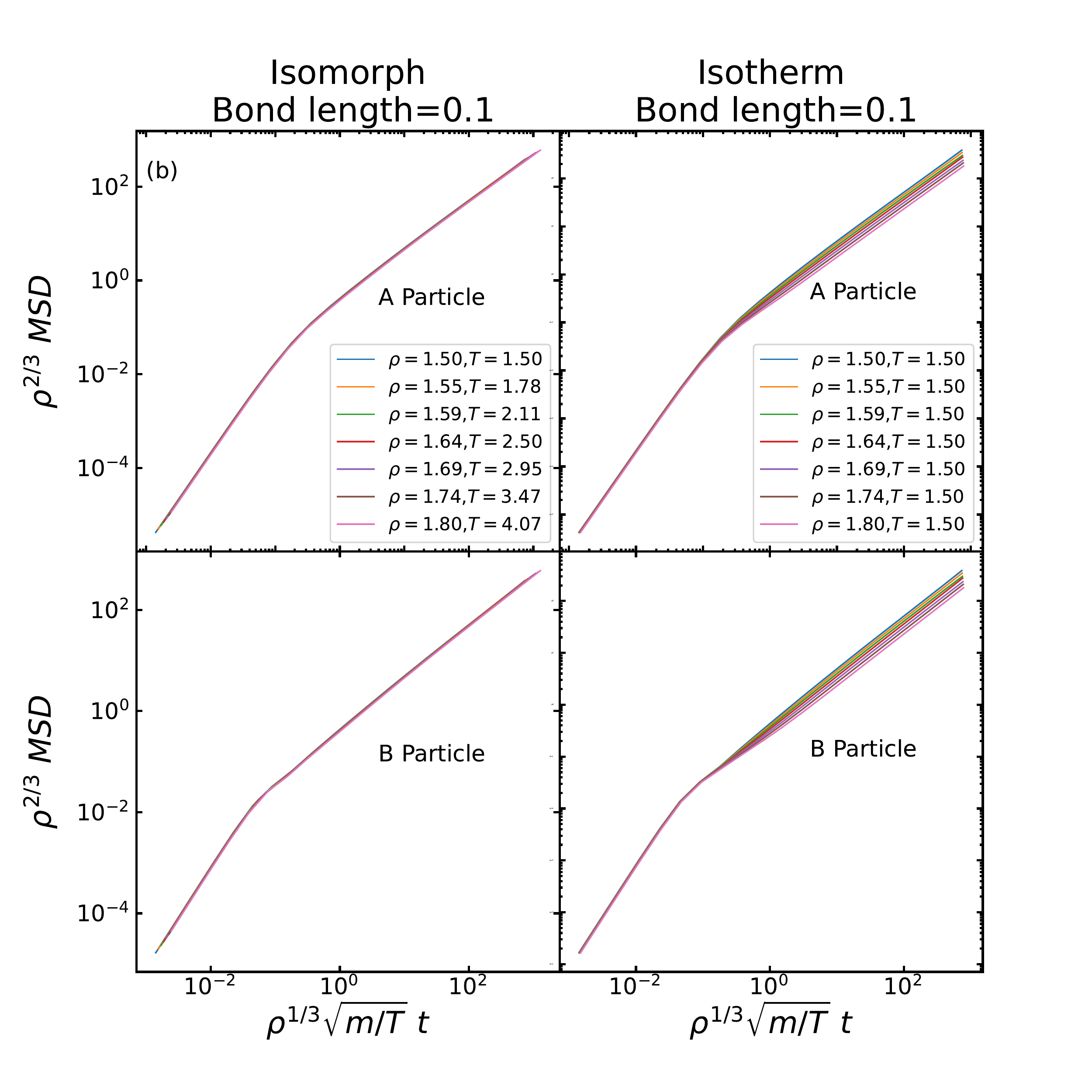}
	\includegraphics[width=7.5cm]{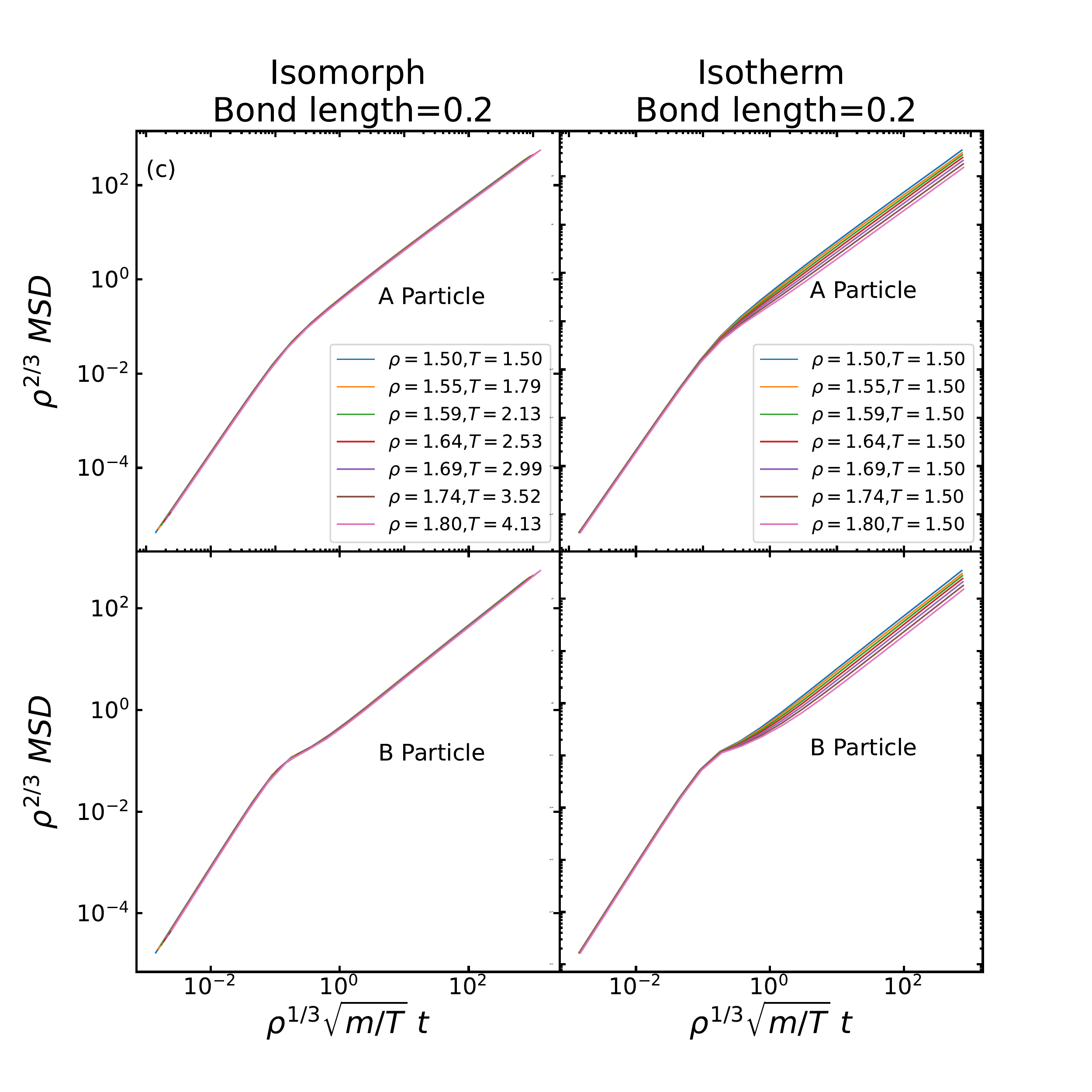}
	\includegraphics[width=7.5cm]{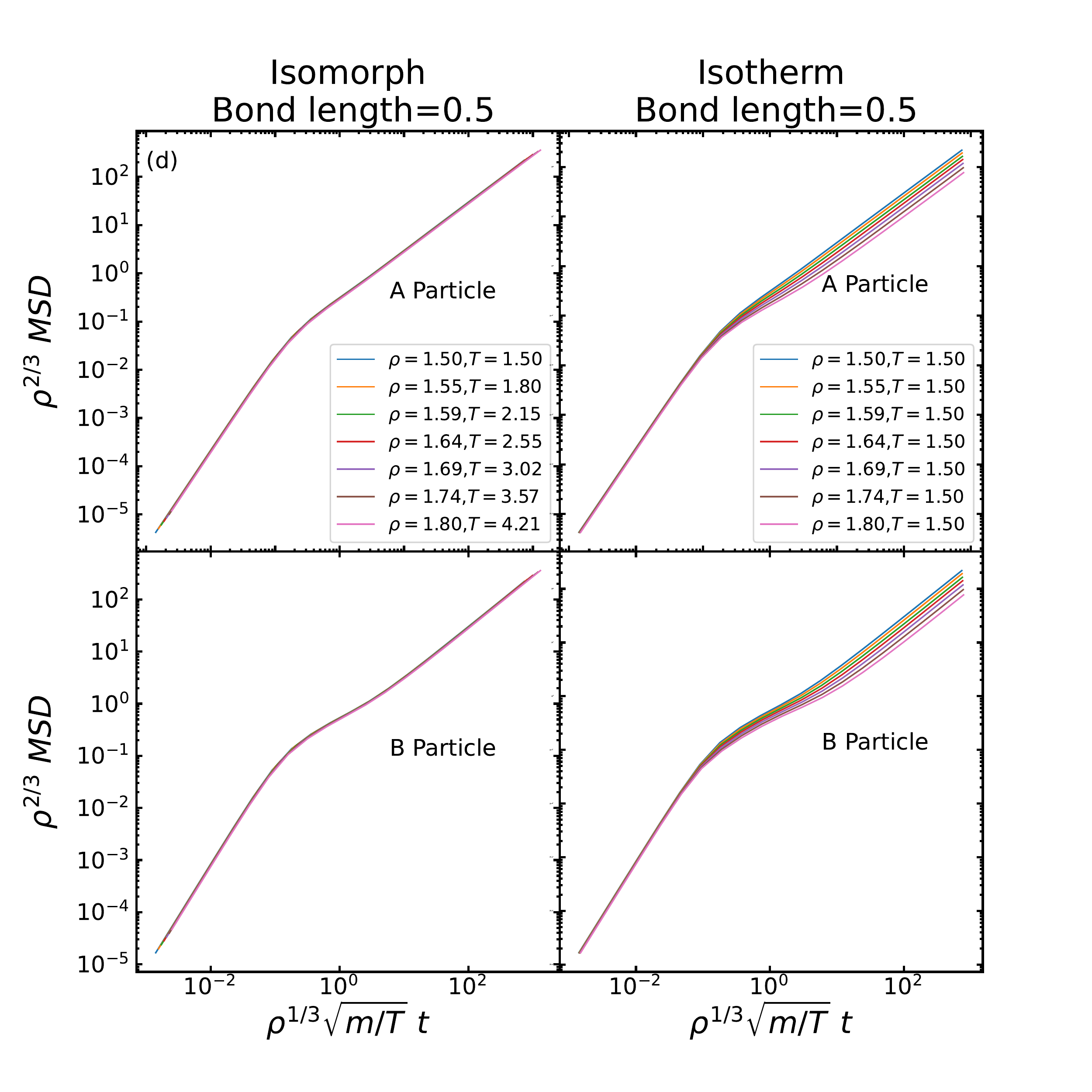}
	\caption{(\textbf{a}) The A and B reduced-unit mean-square displacement (MSD) along the isomorph and the isotherm for bond length $0.05$. There is invariance along the isomorph for both the A and B reduced MSD, but not along the corresponding isotherm.
	(\textbf{b}--\textbf{d}) show similar plots for bond lengths 0.1, 0.2, and 0.5. There is invariance along the isomorphs for both MSDs, but not along the corresponding isochores.}
	\label{fig4}  	
\end{figure}

\begin{figure}[H]
  \centering  
  \includegraphics[width=7.5cm]{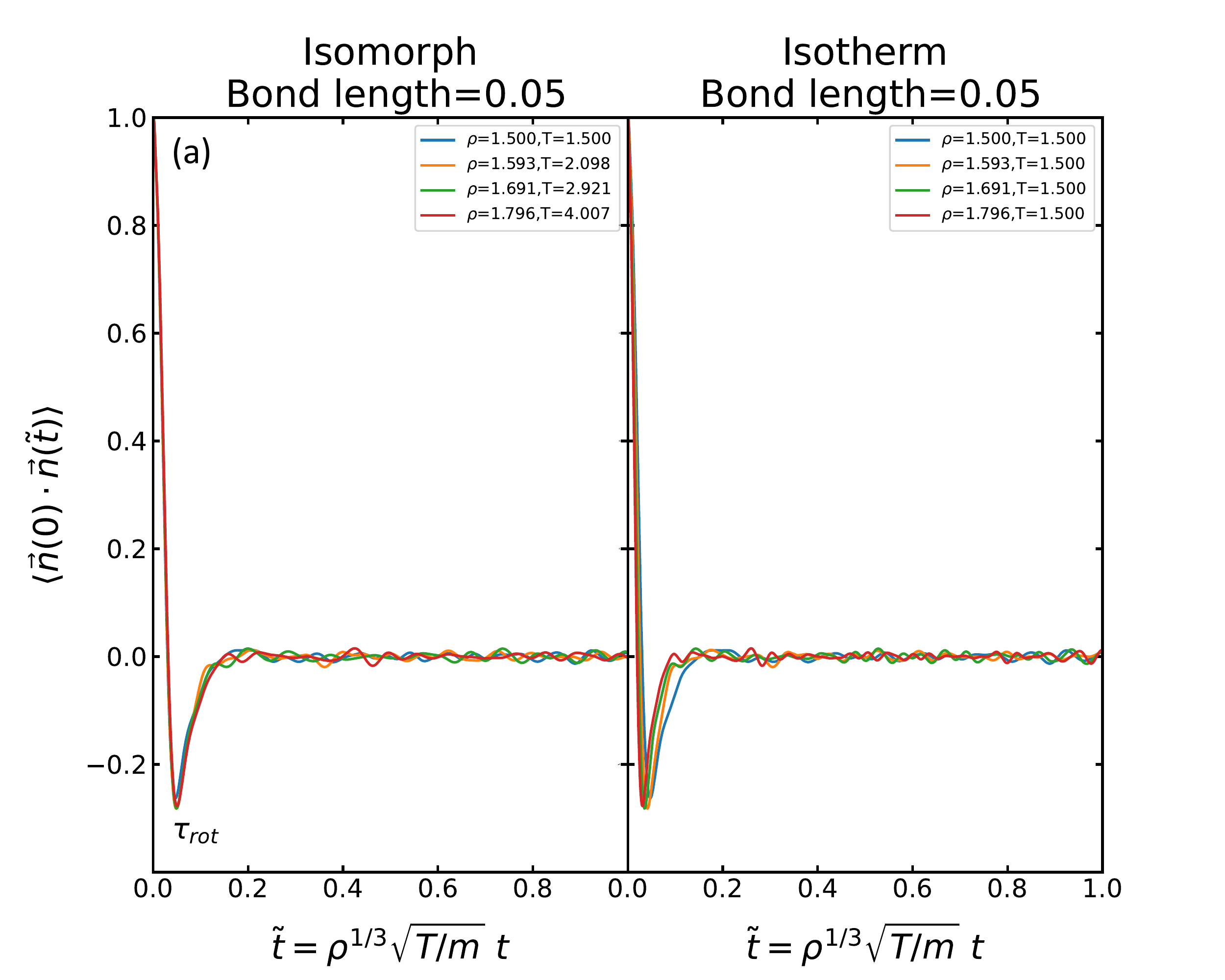}
  \includegraphics[width=7.5cm]{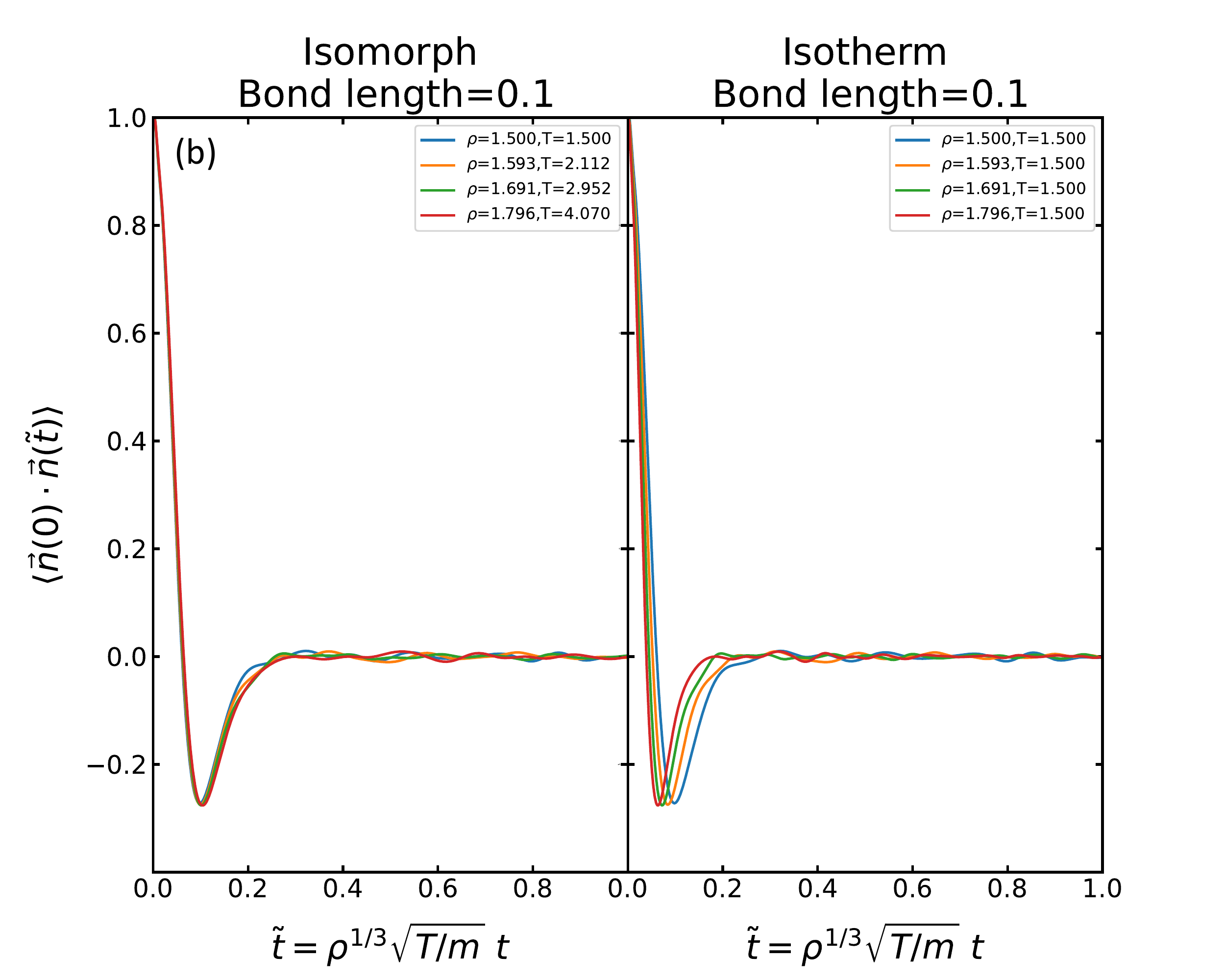}
  \includegraphics[width=7.5cm]{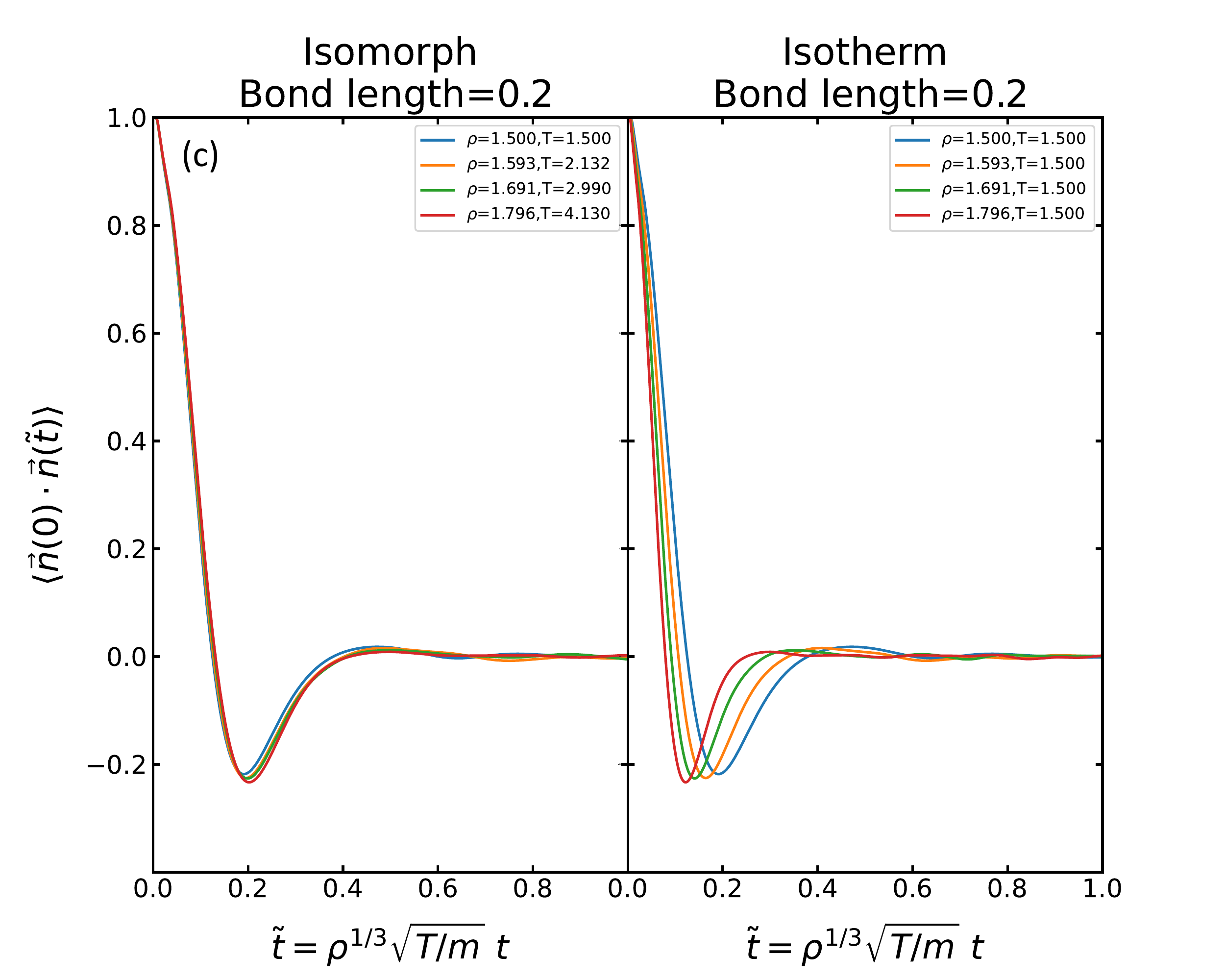}
  \includegraphics[width=7.5cm]{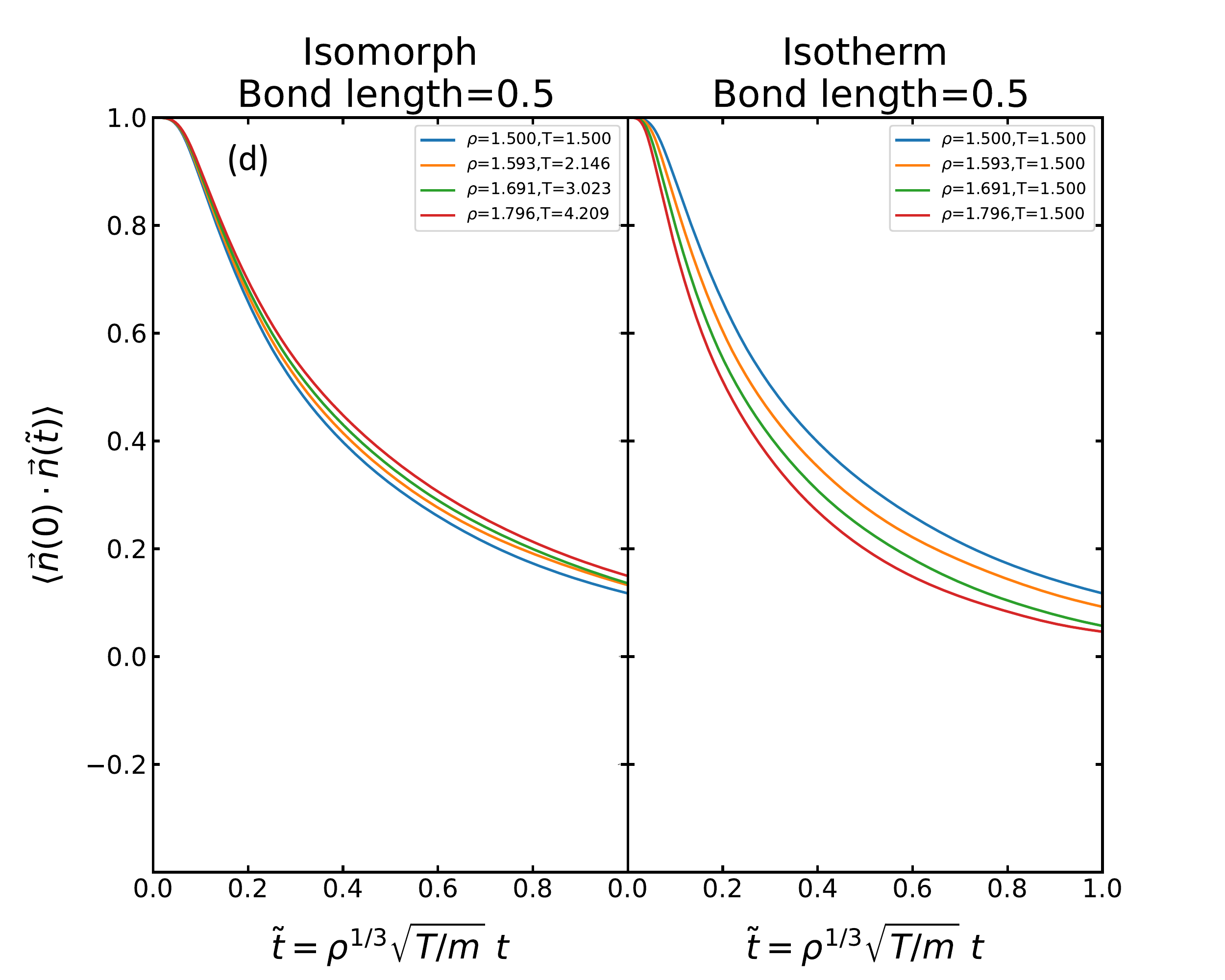}\vspace{5pt}
  \caption{ \textbf{a}--\textbf{d} The rotational time-autocorrelation function (RAC) along the isomorphs and the isotherm for the four different bond lengths. As the bond length increases, the decay to zero becomes significantly slower. The largest bond length (0.5) behaves differently from the others by not going below zero, confirming the change of physics suggested by Fig. 1c. }
 \label{fig5}
\end{figure}

\Fig{fig4} shows data for the (reduced) time dependence of the reduced mean-square displacement (MSD)~\cite{tildesley} of the A and B particles in a log-log plot. The data are isomorph invariant to a good approximation, but not invariant along the isotherm except in the short-time ballistic regime where the invariance is a rigorous consequence of the use of reduced units, which implies unity thermal velocity. Focusing on the behavior along the isomorphs, we note that the A particle motion is pretty similar for all bond lengths. This is consistent with the above picture according to which the A particles to a significant extent behave as if the B particles were not present. In the ballistic regime, the B particles move faster than the A particles because of their lower mass. At long times, the A and B particles are constrained to follow each other, resulting in the same long-time MSD for all bond lengths (which, of course, also applies along the isotherms). An interesting feature appears at intermediate times for the B particle MSD, which for small bond lengths have a slight kink that at larger bond lengths develops into a hint of a plateau. We interpret this as an effect of the fact that the fast ballistic motion of the B particles around the A particles eventually ``saturates''. This is confirmed in Figure~\ref{fig5} discussed next, showing that the rotational time-autocorrelation function for short bond lengths has a negative minimum in this range of the reduced times (around 0.1).

\Fig{fig5} shows data for the rotational time-autocorrelation function (RAC) defined as the autocorrelation of the unit vector from particle A to particle B, $\bf{n}$, plotted as a function of the reduced time. Because of the normalization, this quantity always starts in unity at time zero. The upper row shows data for the isotherms, the lower for the isomorphs. There is little difference and good collapse in both cases, except for the largest bond length 0.5 for which the data are significantly less invariant along the isotherm. At this bond length there is also a change of behavior in the sense that the RAC never becomes negative as it does at intermediate times for the lower bond lengths. A negative value of the RAC signals a more than 90 degree average rotation of the molecule. This finding for bond length 0.5 is consistent with the above observation that a change of the physics appears to set in at bond lengths 0.3--0.4. It is important to note that, in contrast to the RDF and the MSD, the RAC is not predicted to be isomorph invariant because the moment of inertia is not isomorph invariant (the bond length is fixed and not scaled with the density by the factor $\rho^{-1/3}$ required for a constant reduced-unit moment of inertia). 

\section{Results for the Plastic-Crystal Phase}

\begin{figure}[H]\centering
  	\includegraphics[width=5.2cm]{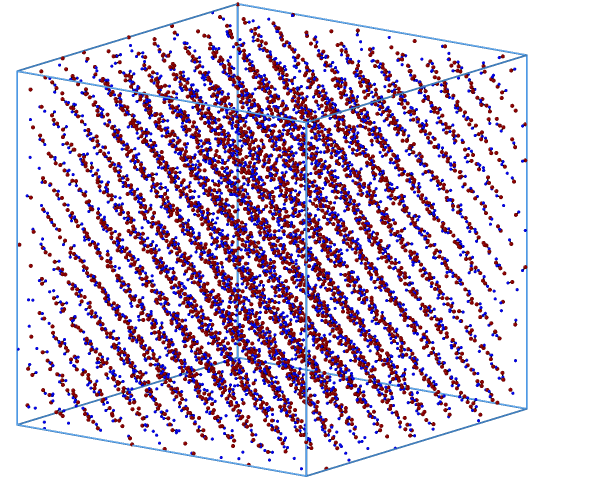}
  	\caption{Snapshot of a plastic crystal configuration at the reference state point $(\rho,T)=(2.2,0.5)$. The bond length between particles A (red) and B (blue) is in this case $0.2$.}
  	\label{fig6} 
\end{figure}

We proceed to report results for the plastic-crystal phase for the bond lengths 0.05, 0.1, 0.2, and 0.3 (for bond lengths larger than 0.3 the system was not crystalline at the chosen reference state point of density 2.2 and temperature 0.5). Table \ref{tab2} gives the pressure, density-scaling exponent, and virial potential-energy correlation coefficient for the four bond lengths at the reference state point. A typical configuration is shown in Figure~\ref{fig6}. We clearly see that the system is ordered (compare Figure~\ref{fig1}a). A closer inspection reveals that the order is not perfect, however, because the bond directions are disordered. This is a typical signal of a plastic crystal.  At a lower temperature, there must be a phase transition to a perfectly ordered phase in which the bond orientations are also ordered, but we did not investigate that transition.

\begin{table}[H]
	\caption{\label{tab2} Thermodynamic parameters at the plastic-crystal reference state point, giving density, temperature, and pressure, as well as the density-scaling exponent $\gamma$ and the virial potential-energy correlation coefficient $R$.}
	\setlength{\tabcolsep}{3.6mm}\begin{tabular}{cccccc}
		\toprule
		\textbf{Bond Length [}$\boldsymbol{\sigma}_{\textbf{\emph{AA}}}$\textbf{]} & $\boldsymbol{\rho}$ \textbf{[}$\textbf{1/}\boldsymbol{\sigma}_{\textbf{\emph{AA}}}^\textbf{3}$\textbf{]} & $\textbf{\emph{T}}$ \textbf{[}$\boldsymbol{\varepsilon}_{\textbf{\emph{AA}}}\textbf{/$k_B$}$\textbf{]} & $\textbf{\emph{p}}$ \textbf{[}$\boldsymbol{\sigma}_{\textbf{\emph{AA}}}^{\textbf{-3}}\boldsymbol{\varepsilon}_{\textbf{\emph{AA}}}$\textbf{]} & $\boldsymbol{\gamma}$ & $\textbf{\emph{R}}$ \\
		0.05 & 2.2 & 0.5 & 0.50 & 5.81 &  0.997 \\
		0.10 & 2.2 & 0.5 & 1.17 & 5.61 &  0.996  \\
		0.20 & 2.2 & 0.5 & 5.19 & 5.25 &  0.992 \\
		0.30 & 2.2 & 0.5 & 15.28 & 5.37 & 0.990\\
	\end{tabular}
\end{table}
\vspace{-12pt}

\begin{figure}[H]
    \centering	\includegraphics[width=5.9cm]{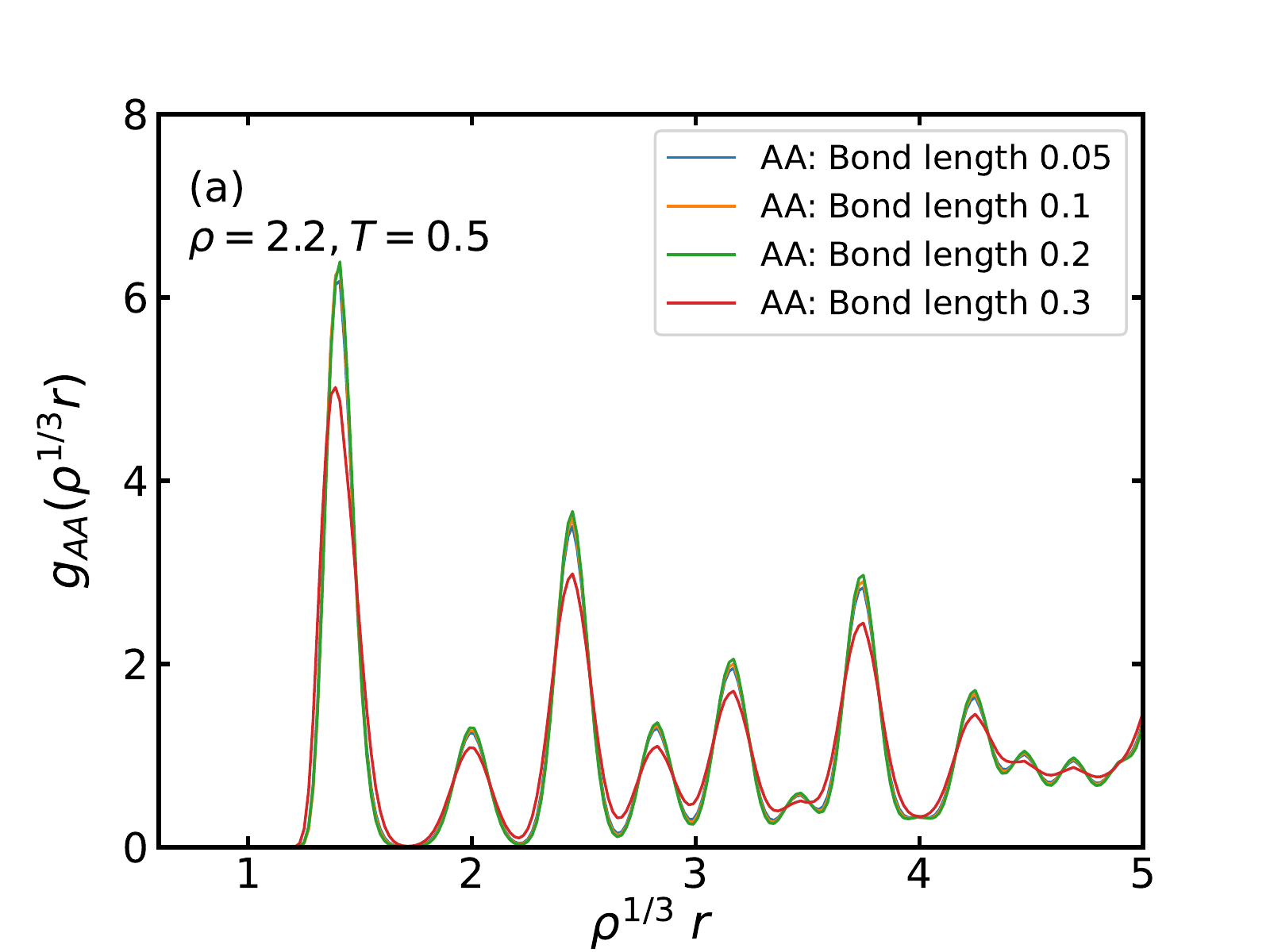}~~~~
    \includegraphics[width=5.9cm]{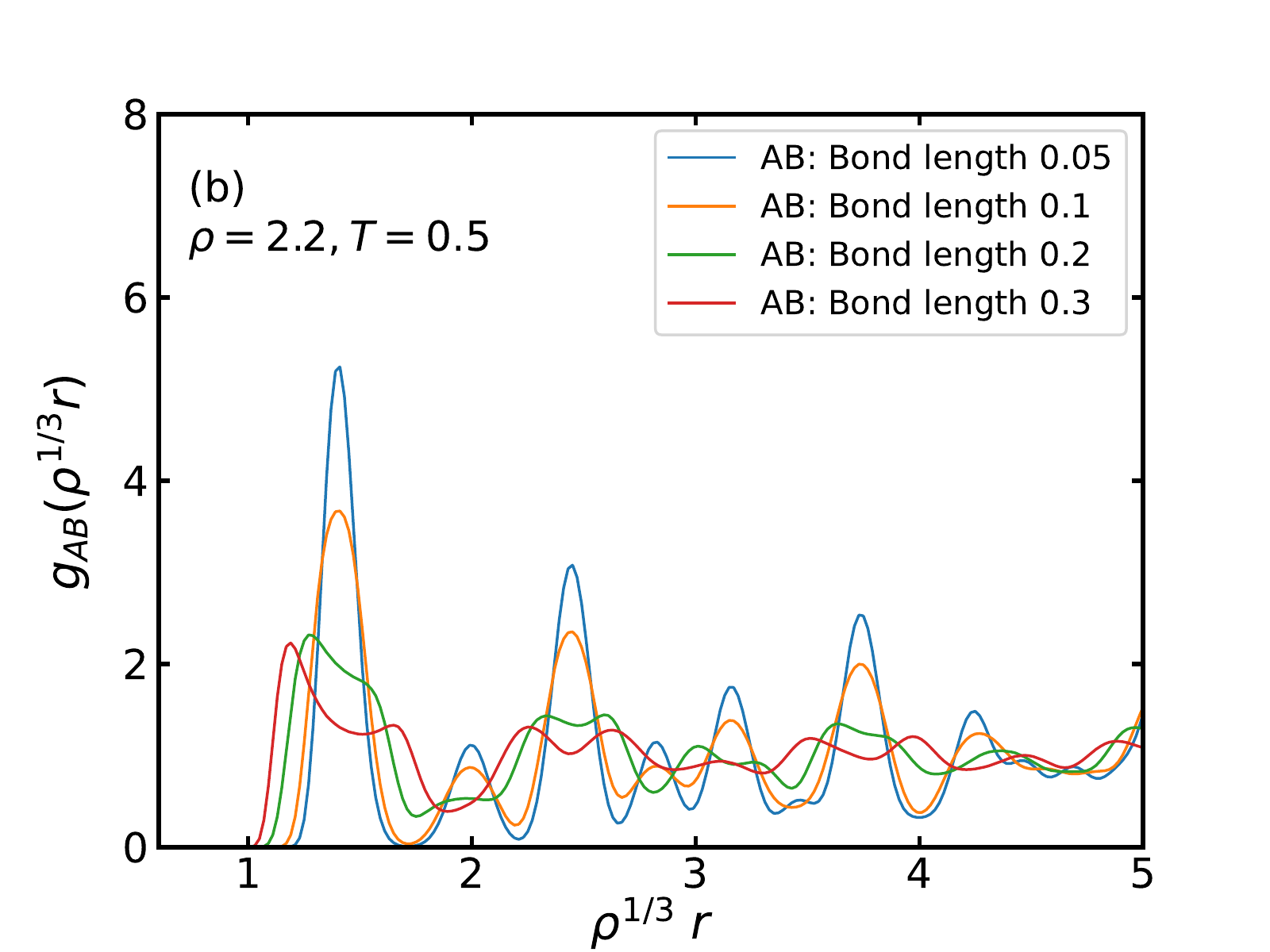}~~~~
  	\includegraphics[width=5.9cm]{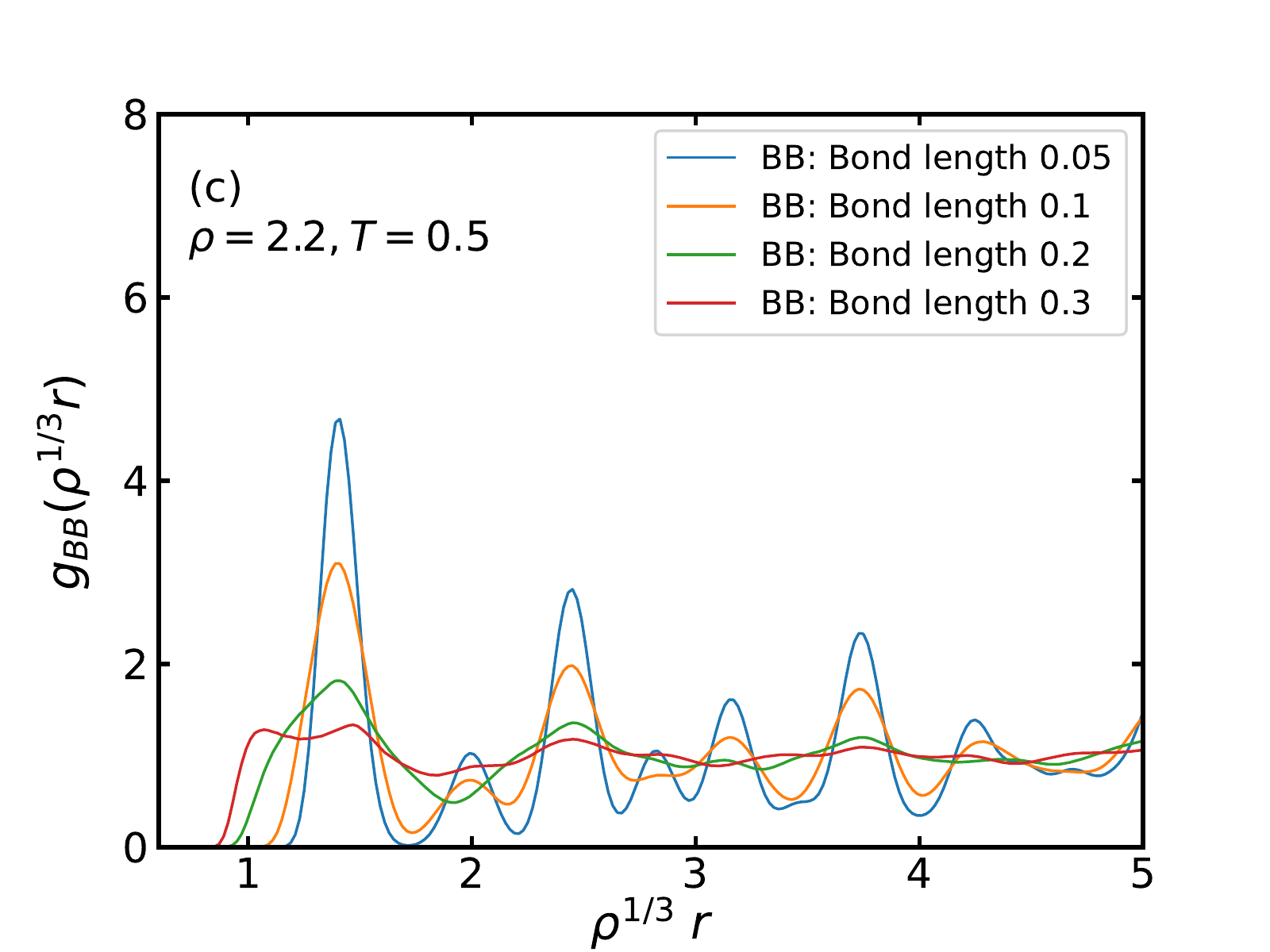}
  	\caption{(\textbf{a}) AA particle RDFs at the reference state point $\rho=2.2, T=0.5$ for each of the bond lengths $0.05, 0.1, 0.2, 0.3$. 
  	(\textbf{b},\textbf{c}) AB and BB particle RDFs, respectively, at the same state point.}
  	\label{fig7} 
\end{figure}

Before proceeding to probe the extent of isomorph invariance, we look at the properties at the reference state point. \Fig{fig7}a shows the AA particle RDFs for the four bond lengths. These show a well-defined order due to the (almost) crystalline ordering of the A particles; the AA particle RDFs are moreover very similar for all bond lengths. In contrast, the AB and BB RDFs vary significantly because of the rotation of the B particles around the A particles. In these cases, the most ordered RDFs are those of the shortest bonds. These observations show that the slaving of the B particles to the A particles (that are well ordered) applies also in the crystalline phase. 

\begin{figure}[H]\centering
  	\includegraphics[width=5.9cm]{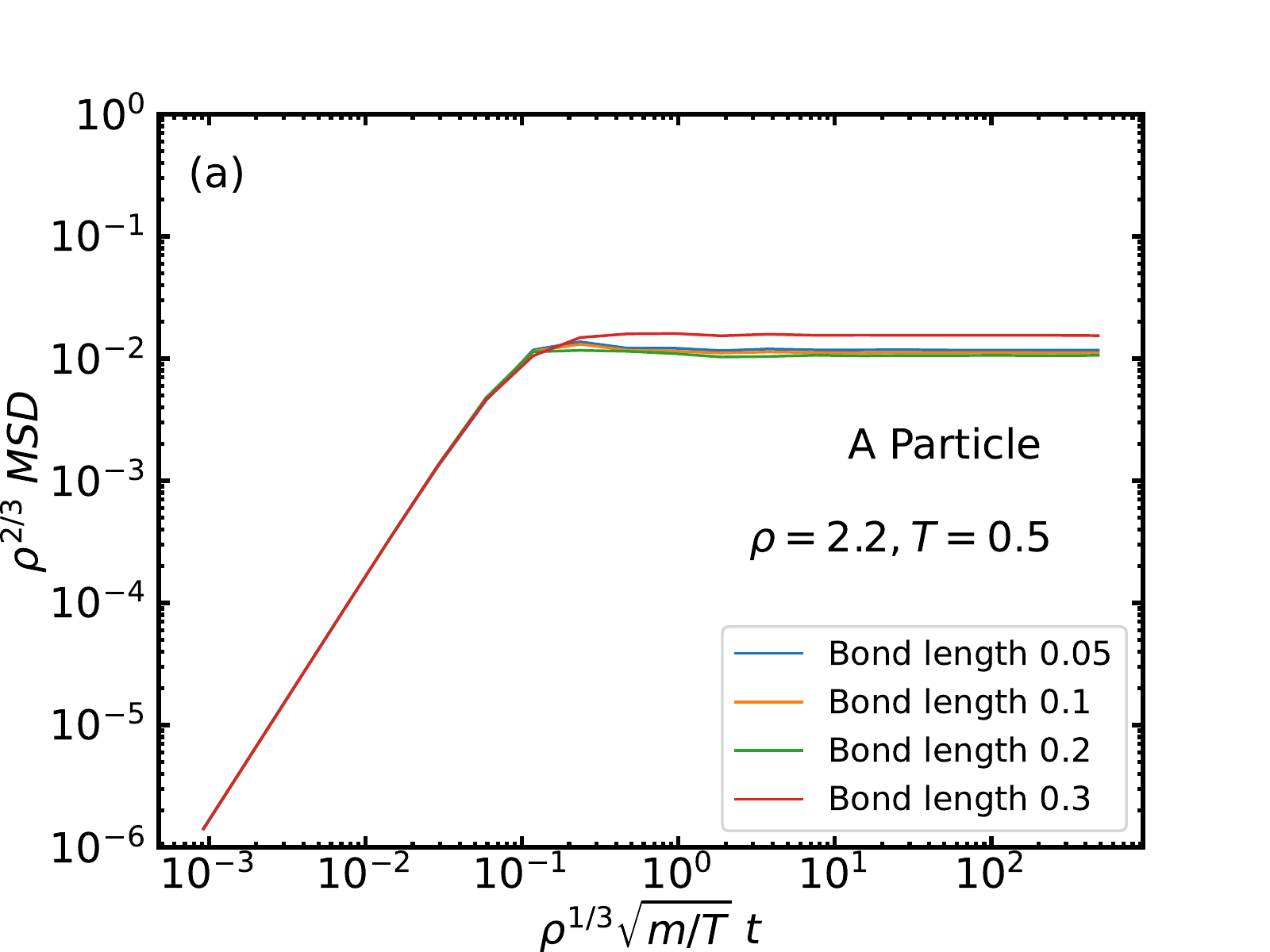}
  	\includegraphics[width=5.9cm]{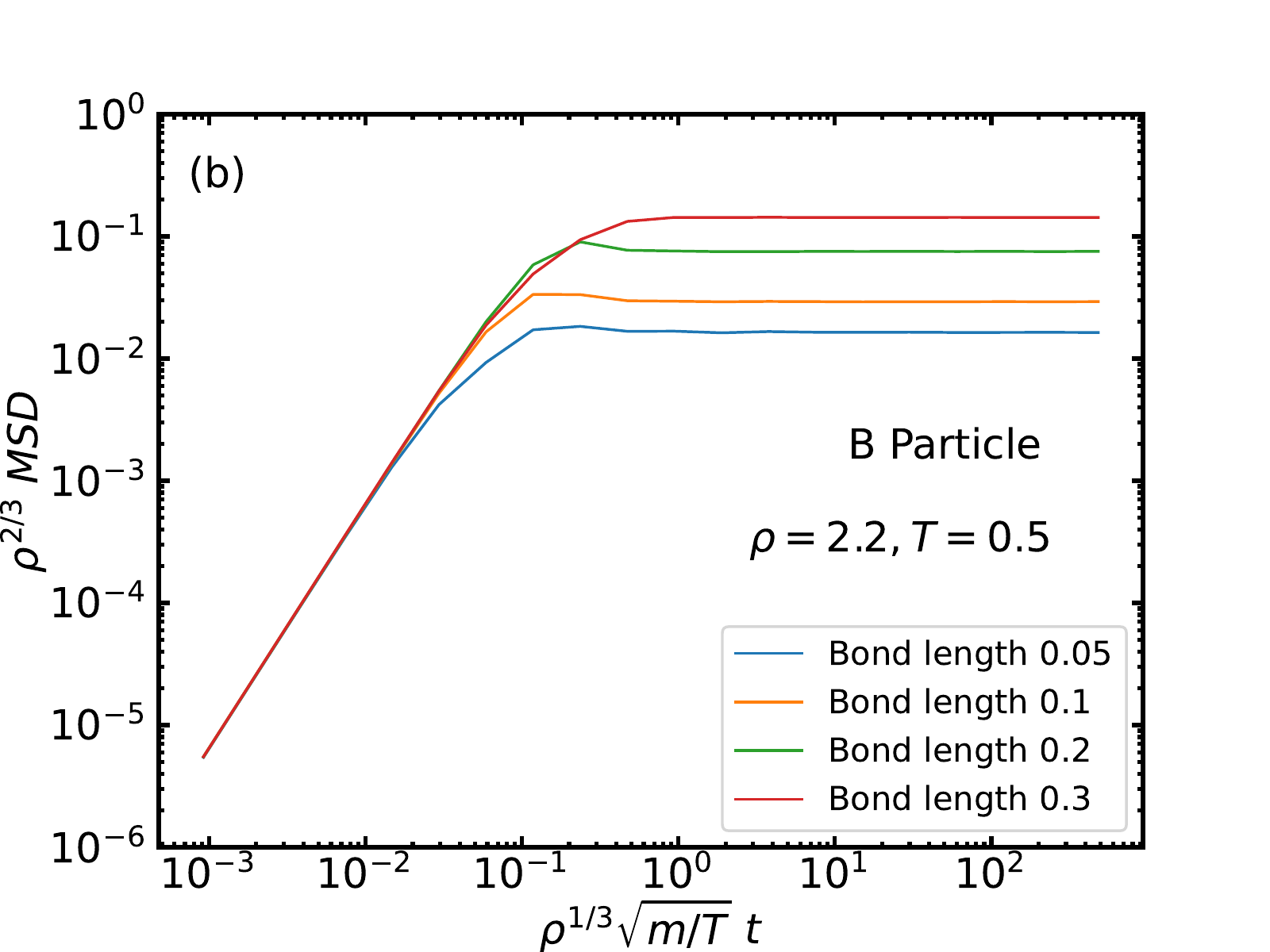}
  	\caption{(\textbf{a}) A particle MSD as a functions of time at the reference state point $(\rho,T)=(2.2,0.5)$ for each of the bond lengths $0.05, 0.1, 0.2, 0.3$. The long-time stabilization to a plateau shows that the system is a solid.
  	(\textbf{b}) B particle MSD at the same state point.}
  	\label{fig8} 
\end{figure}

\Fig{fig8} shows the MSD of the A and B particles as functions of the reduced time. For the A particles the results are very similar for the different bond lengths, although bond length 0.3 deviates from the three smaller ones by having a larger long-time plateau. We have no good explanation for this, but deviations for the longest bond length from the three others are also noted in some of the later figures. The B particle plateaus vary considerably, with larger plateaus observed for larger bond lengths. This is because the B particles rotate around the A particles and a larger bond length gives them more freedom to do so, resulting in a larger long-time plateau.

\begin{figure}[H]\centering
	\includegraphics[width=6.5cm]{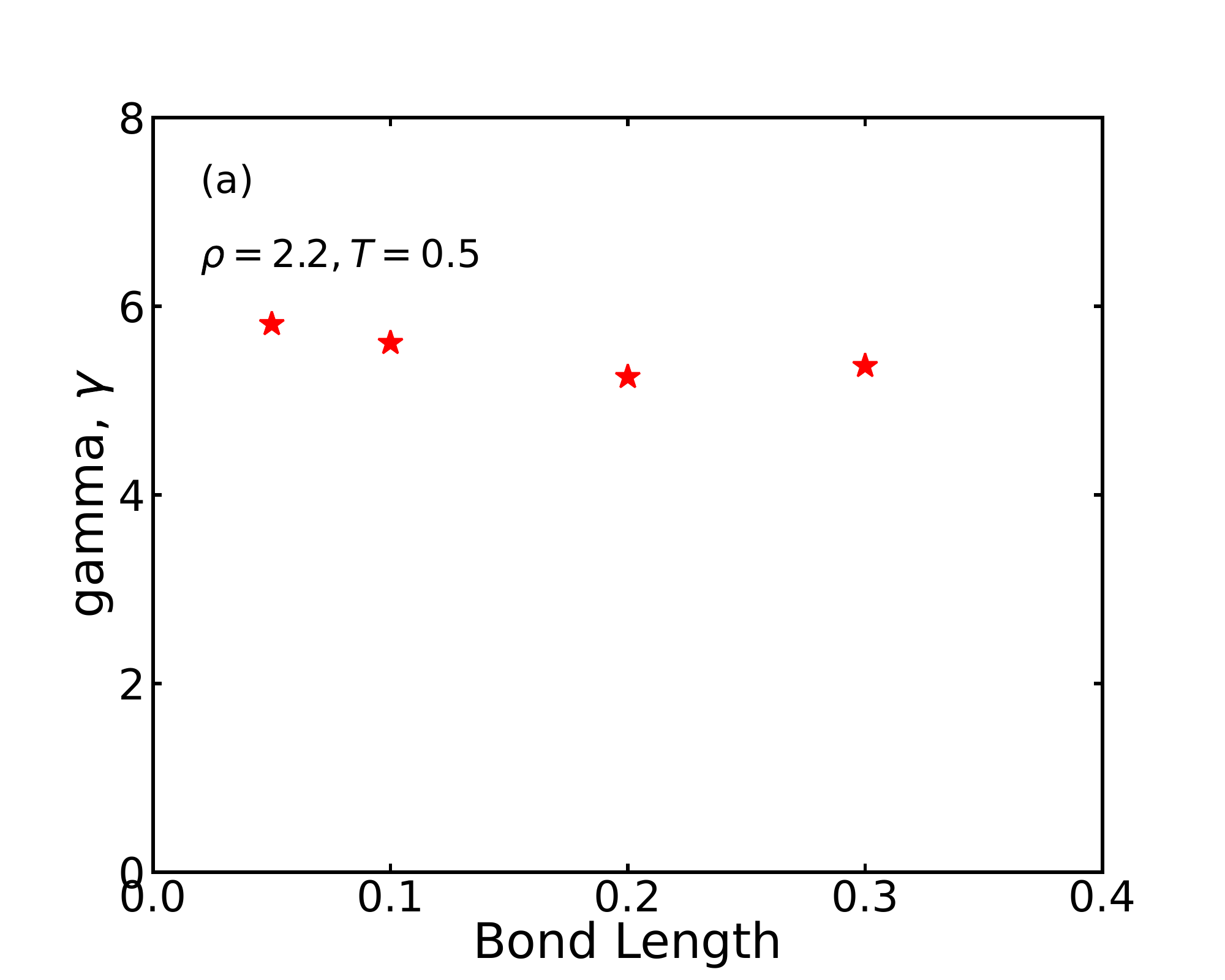} 
	\includegraphics[width=6.5cm]{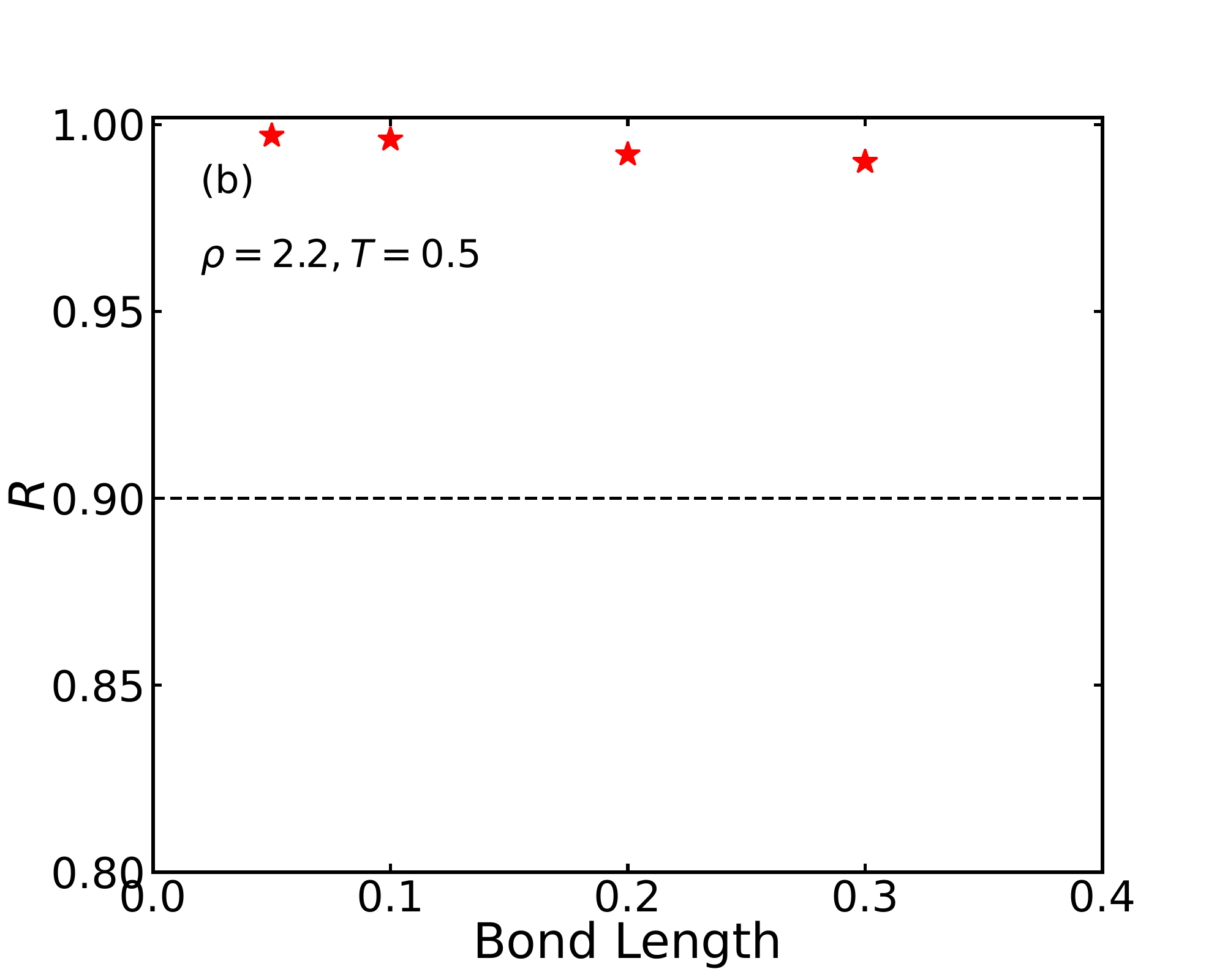}
	\caption{Variation of the density-scaling exponent $\gamma$ (\textbf{a}) and of the virial potential-energy correlation coefficient $R$ (\textbf{b}) at the reference state point plotted as functions of the bond length. Data for a few extra bond lengths have been added.\label{fig9}}
\end{figure}

\Fig{fig9} shows $\gamma$ and $R$ at the reference state point as a function of the bond length. We see that $\gamma$ is almost independent of the bond length and has the typical LJ value between 5 and 6~\cite{I}. The correlation coefficient drops somewhat with increasing bond length, but stays comfortably above 0.9.

\begin{figure}[H]\centering
  	\includegraphics[width=6cm]{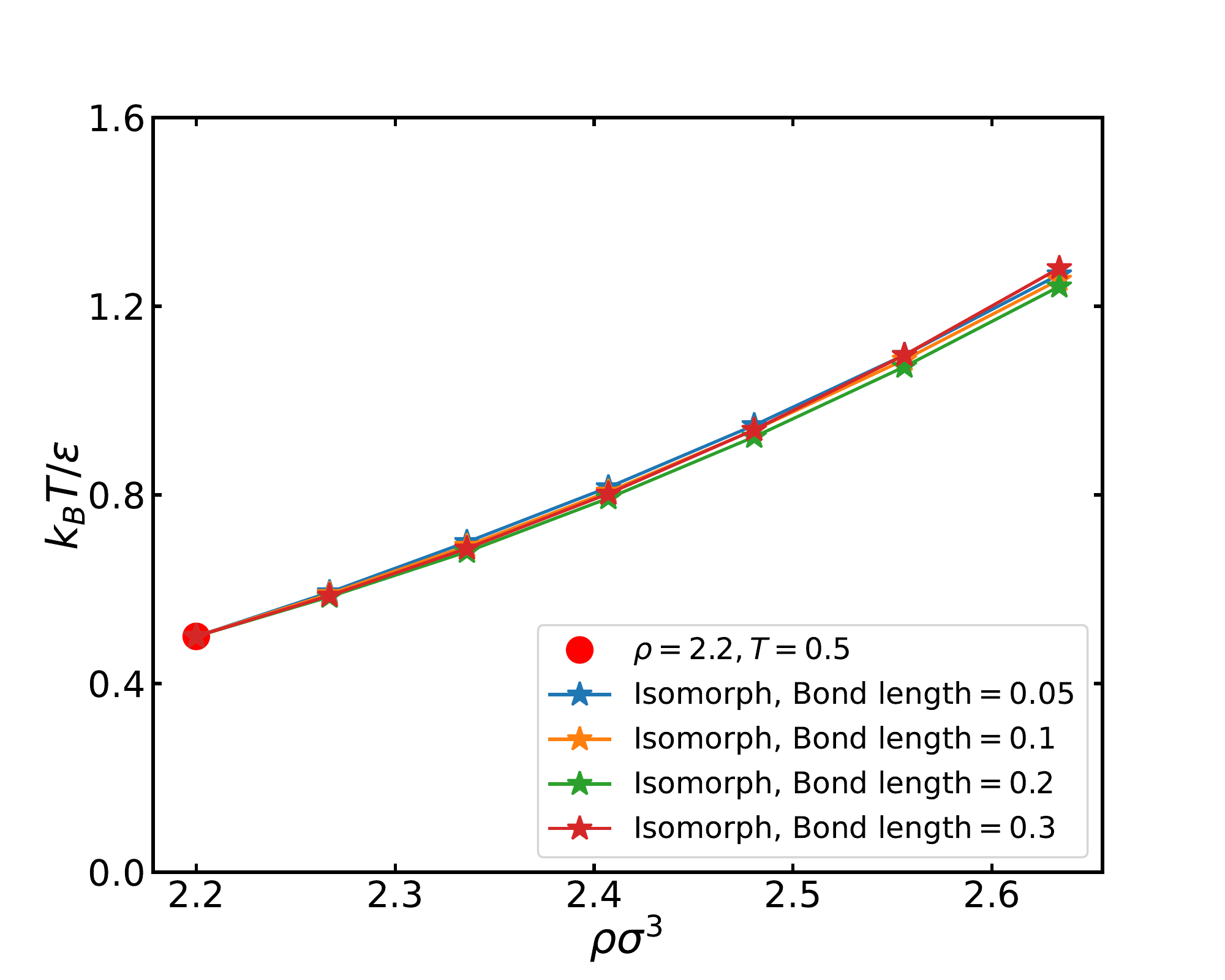}
  \caption{Four isomorphs traced out in the plastic-crystal phase of the ASD thermodynamic phase diagram using the RK4 method with density step size $0.01$. The isomorphs are traced for bond lengths 0.05, 0.1, 0.2, and 0.3, starting from the reference state point marked in red.}
  	\label{fig10} 
\end{figure}

As in the liquid phase, we generated isomorphs from the reference state point by integrating Equation~(\ref{gamma}) using the RK4 algorithm~\cite{att21} for density steps of magnitude 0.01. Additionally, as in the liquid case, we chose to increase the density by 20\% from the reference state-point density. This corresponds to realistic density changes of high-pressure experiments. The four isomorphs are shown in Figure~\ref{fig10}. We see that they are quite similar.

\begin{figure}[H]
    \centering\includegraphics[width=8.9cm]{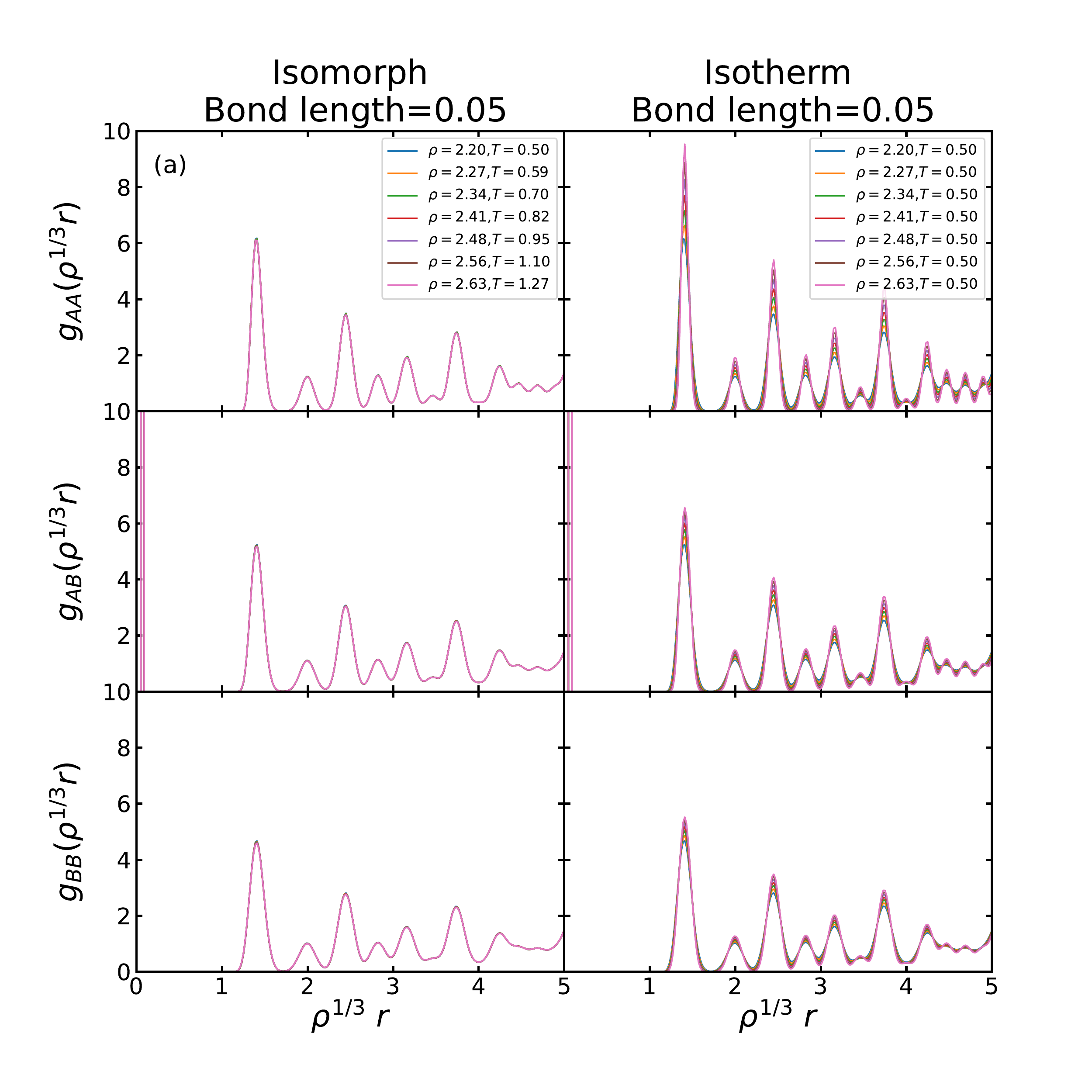}
	\includegraphics[width=8.9cm]{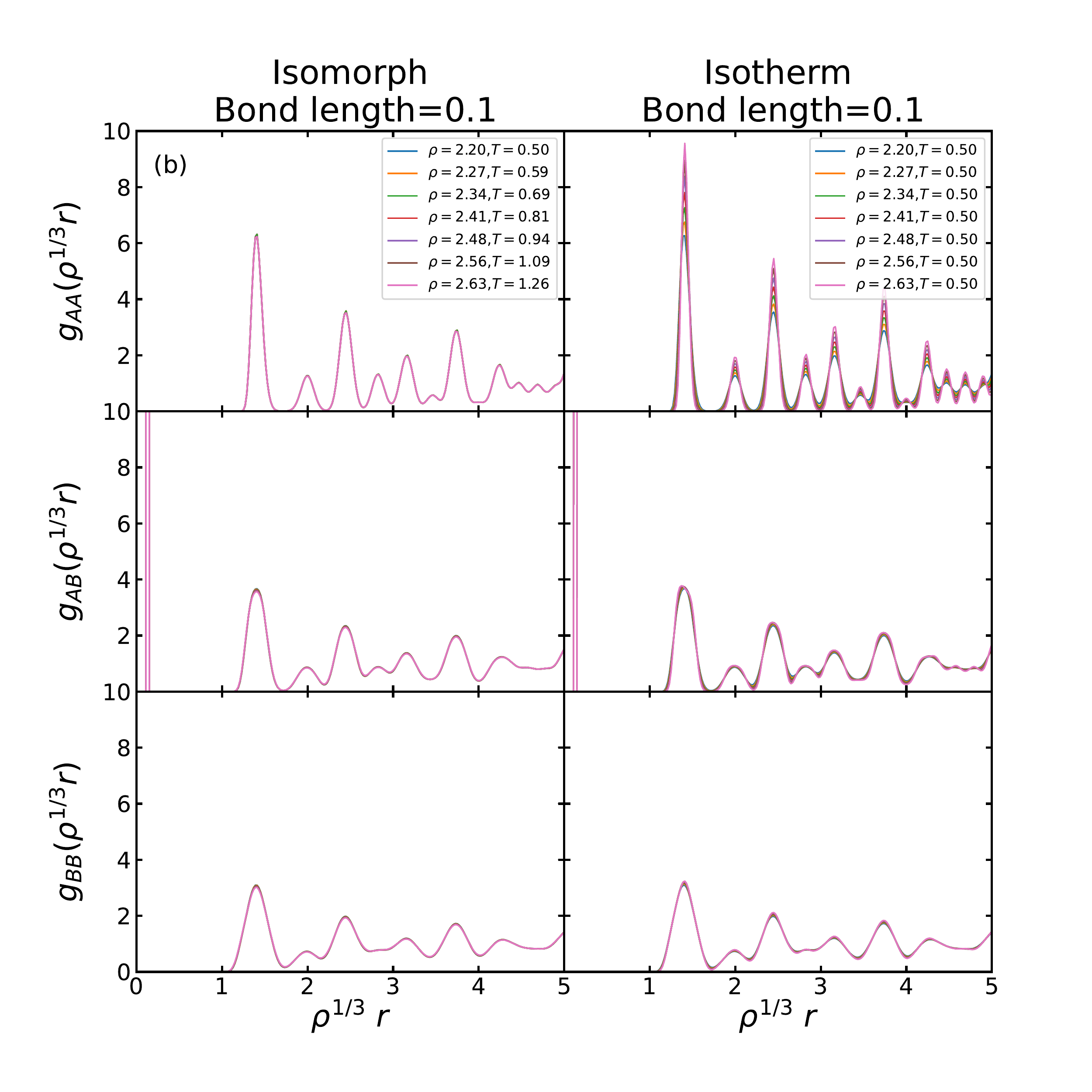}
	\includegraphics[width=8.9cm]{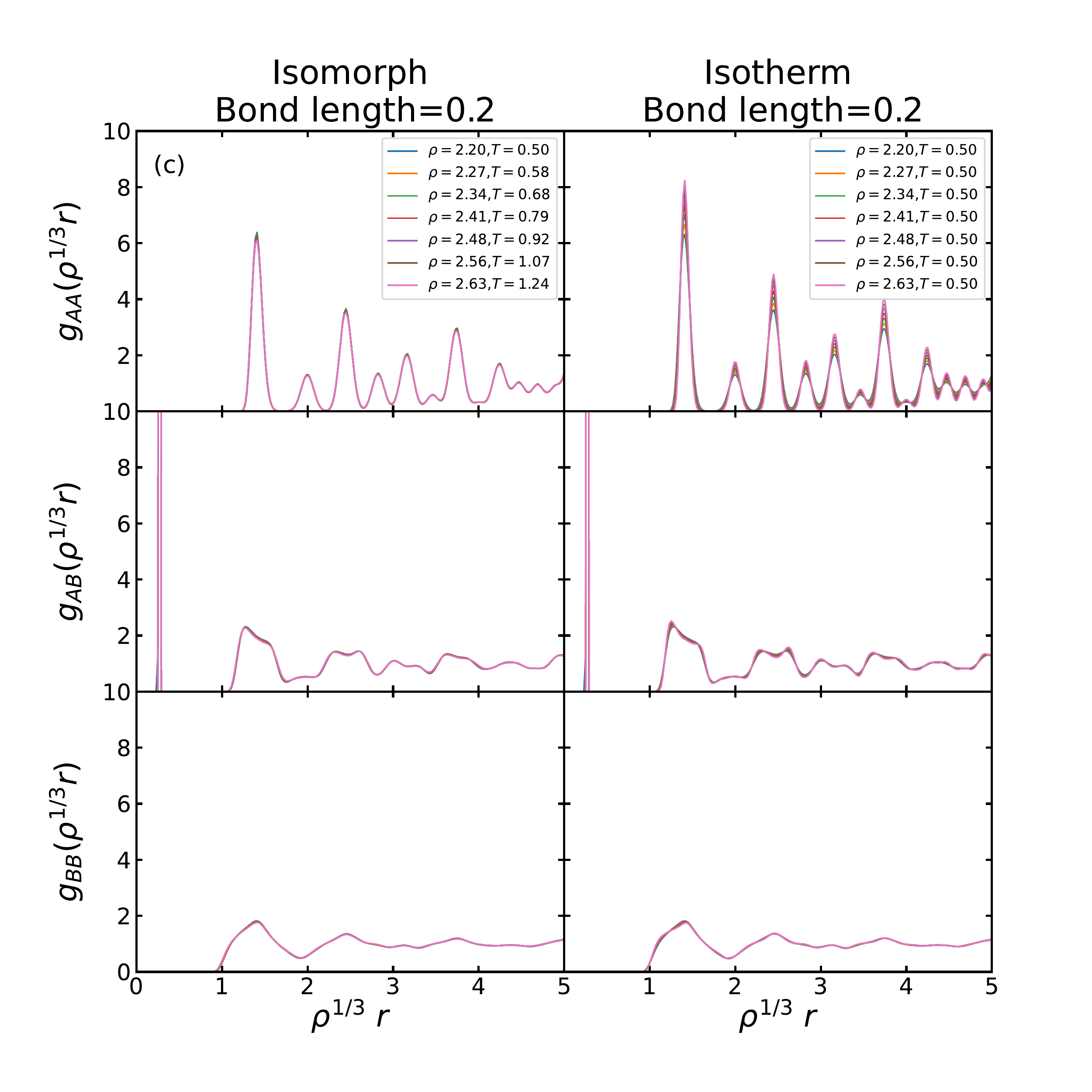}
	\includegraphics[width=8.9cm]{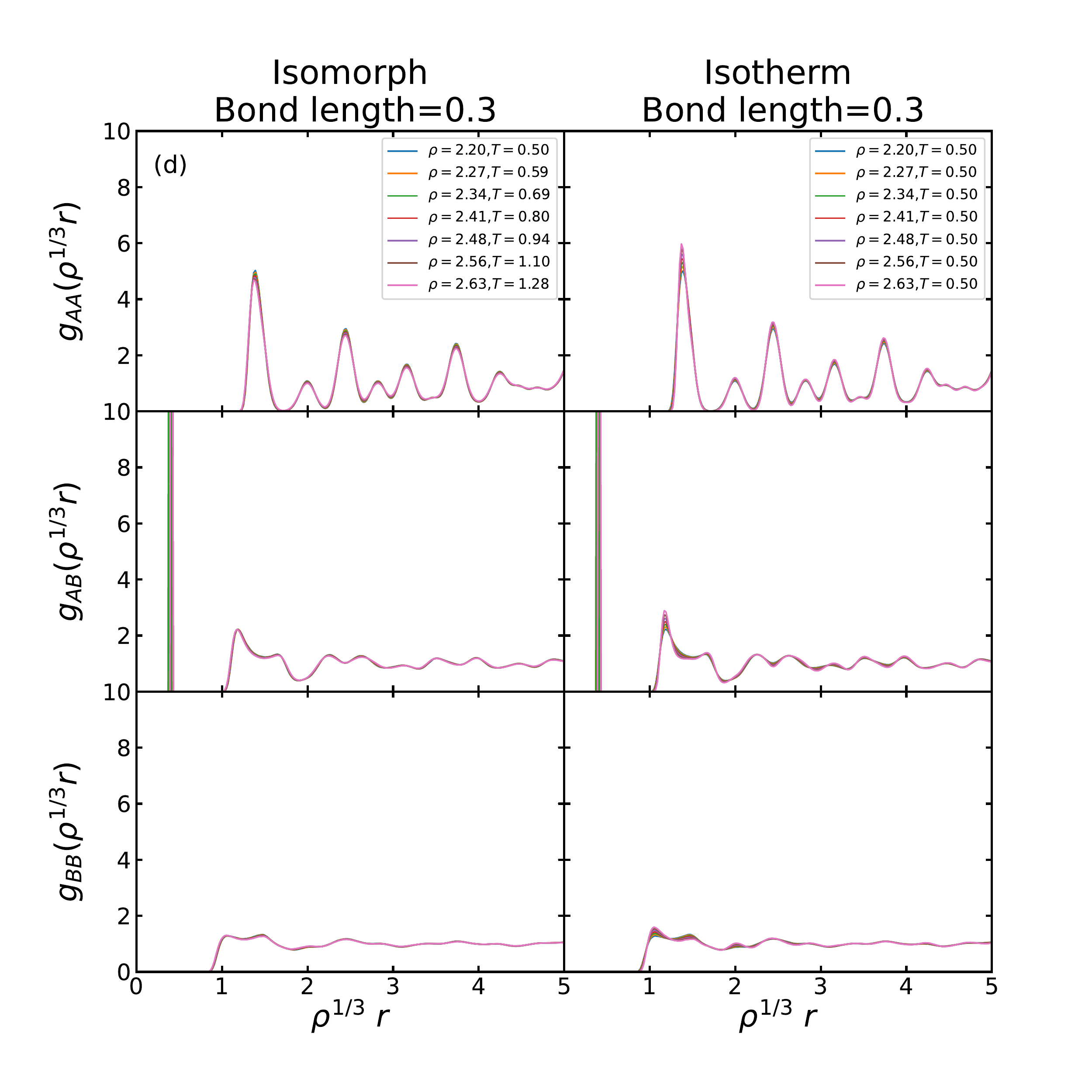}
	\caption{AA, AB and BB RDFs as a function of the reduced pair distance. 
	(\textbf{a}) The RDFs along the isomorph for the bond length 0.05 and, for comparison, along the reference-state-point isotherm of the same (20\%) density variation. We see almost perfect invariance along the isomorph, unlike along the isotherm. The thick vertical line in the AB RDF comes from the fixed bond length, which in reduced units varies with density. 
	(\textbf{b}--\textbf{d}) show similar plots for bond lengths 0.1, 0.2, and 0.3. There is a very good isomorph invariance of all three RDFs in comparison to their isotherm variation. The first peak of the BB RDF gets lower as the bond length increases, which reflects an increased spread of the B particle positions relative to each other. At the same time the AB RDFs also decrease. }
  	\label{fig11}
\end{figure}

\Fig{fig11} shows the reduced RDFs along the isomorphs and, for comparison, along reference-temperature isotherms of the same density variation. For the smallest bond length (0.05), we see a very good collapse of all three RDFs along the isomorph, but not along the isotherm. Moreover, the three RDFs are quite similar, which is a consequence of the fact that all B particles are constrained to be very close to an A particle. For larger bond lengths we see in all cases a very good isomorph collapse and note that, with increasing bond length, the AB and BB particle RDFs become increasingly smeared out compared to the AA particle RDF. The same was seen in the liquid phase (Figure~\ref{fig3}), and as for the liquid phase we interpret this as deriving from rotations of the molecules. In particular, it confirms that the crystals are plastic. Interestingly, the AA particle RDF becomes more invariant along the isotherm as the bond length is increased, but for all bond lengths this quantity is less isotherm than isomorph invariant.

\begin{figure}[H]
    \centering\includegraphics[width=8.9cm]{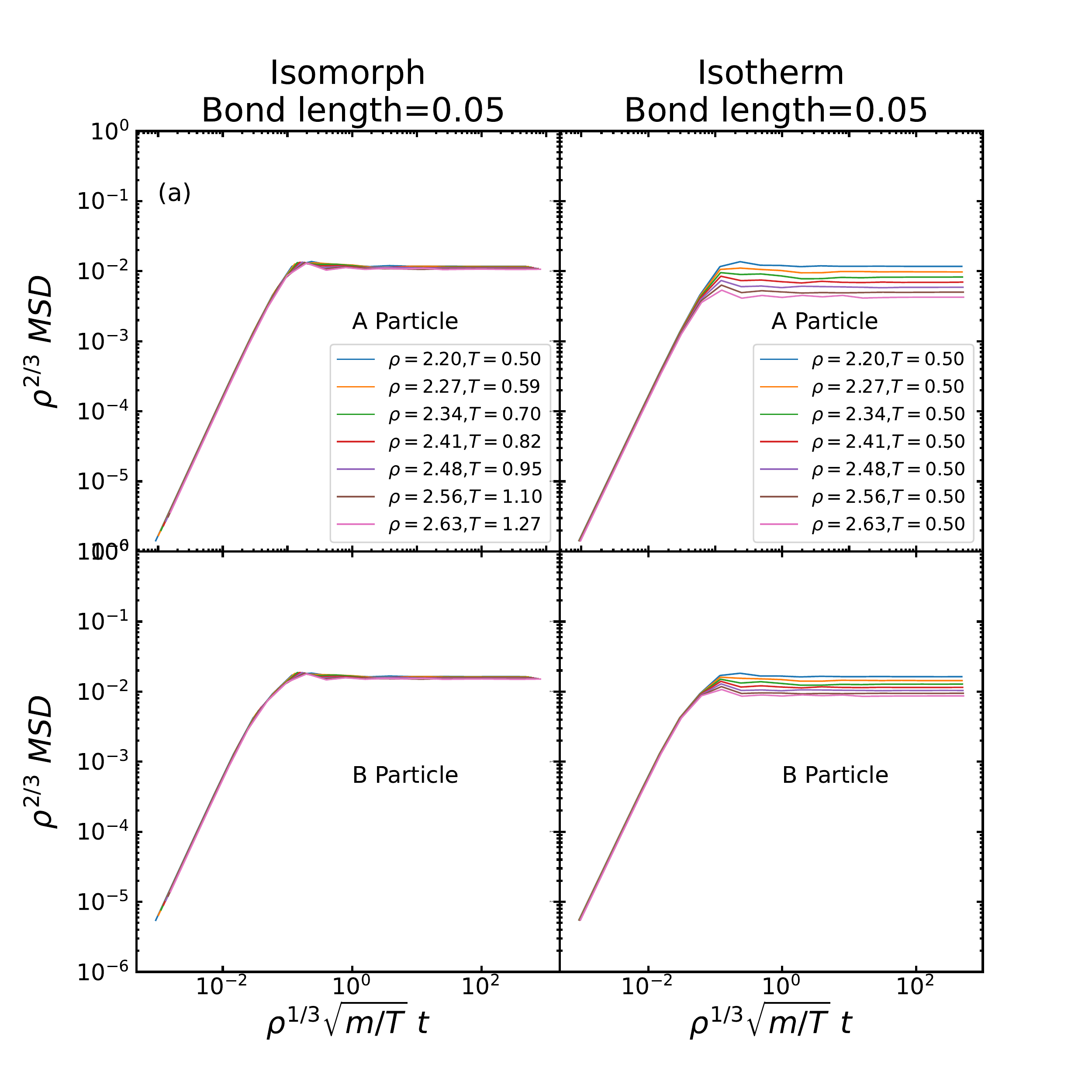}
	\includegraphics[width=8.9cm]{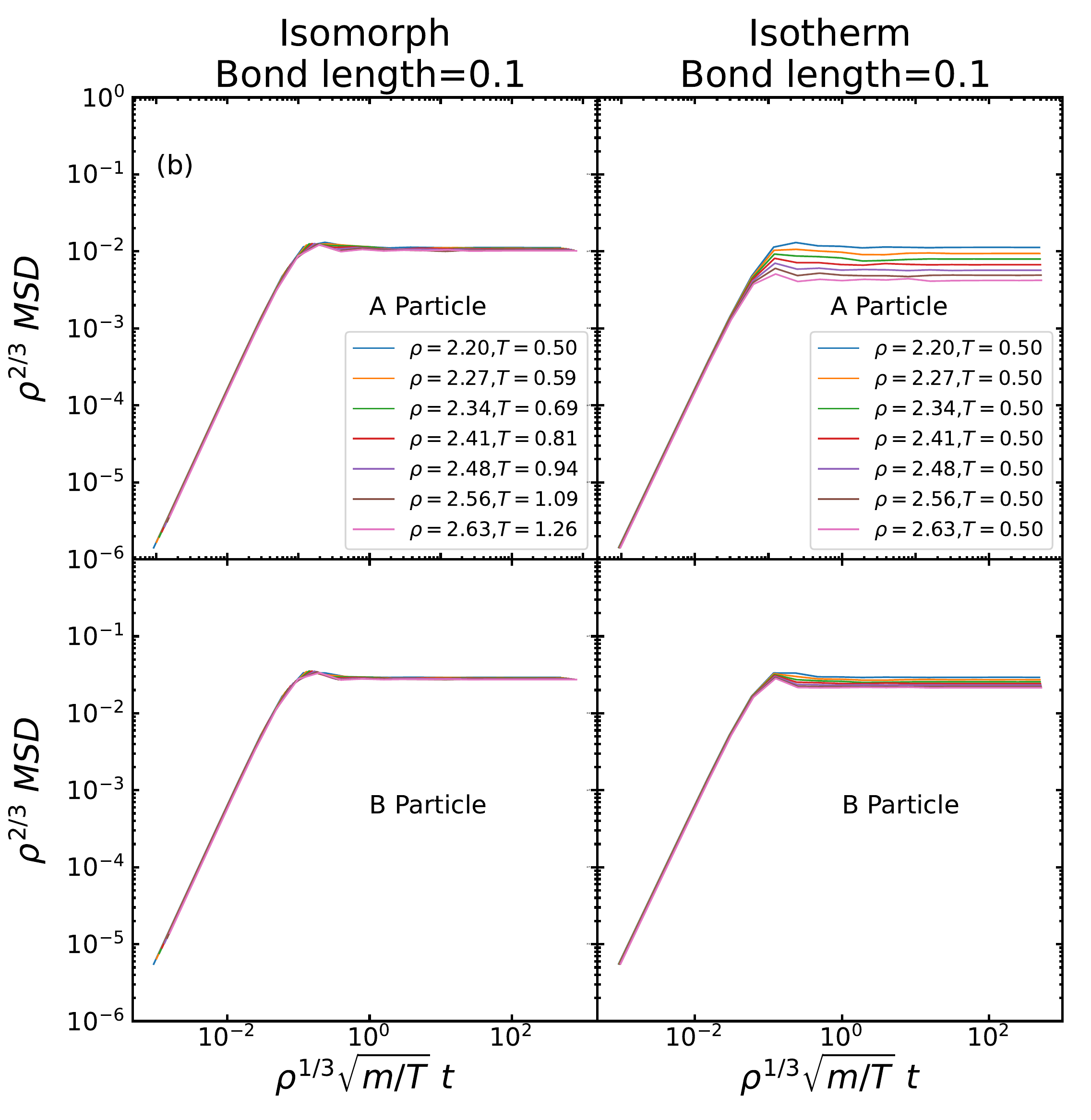}
	\includegraphics[width=8.9cm]{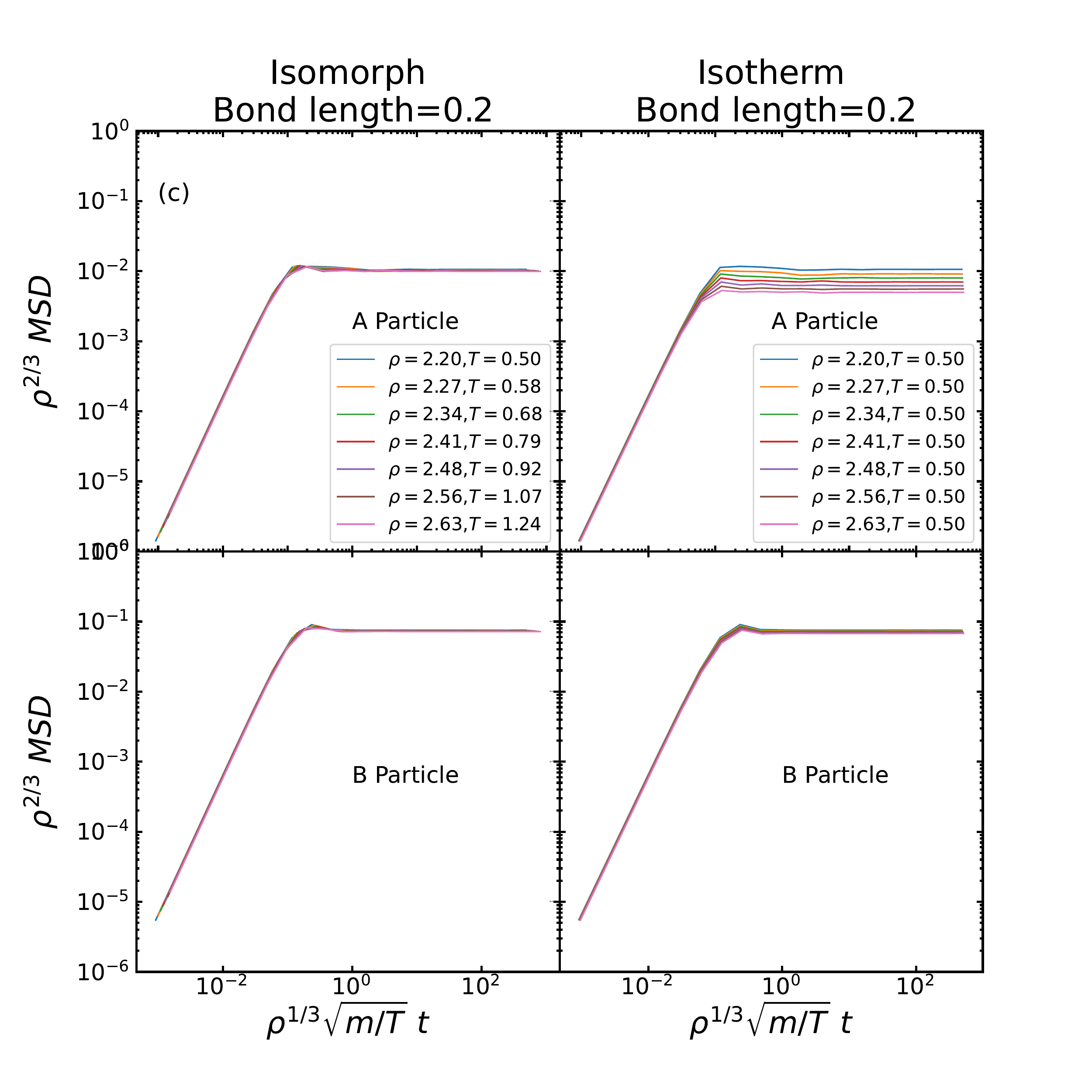}
	\includegraphics[width=8.9cm]{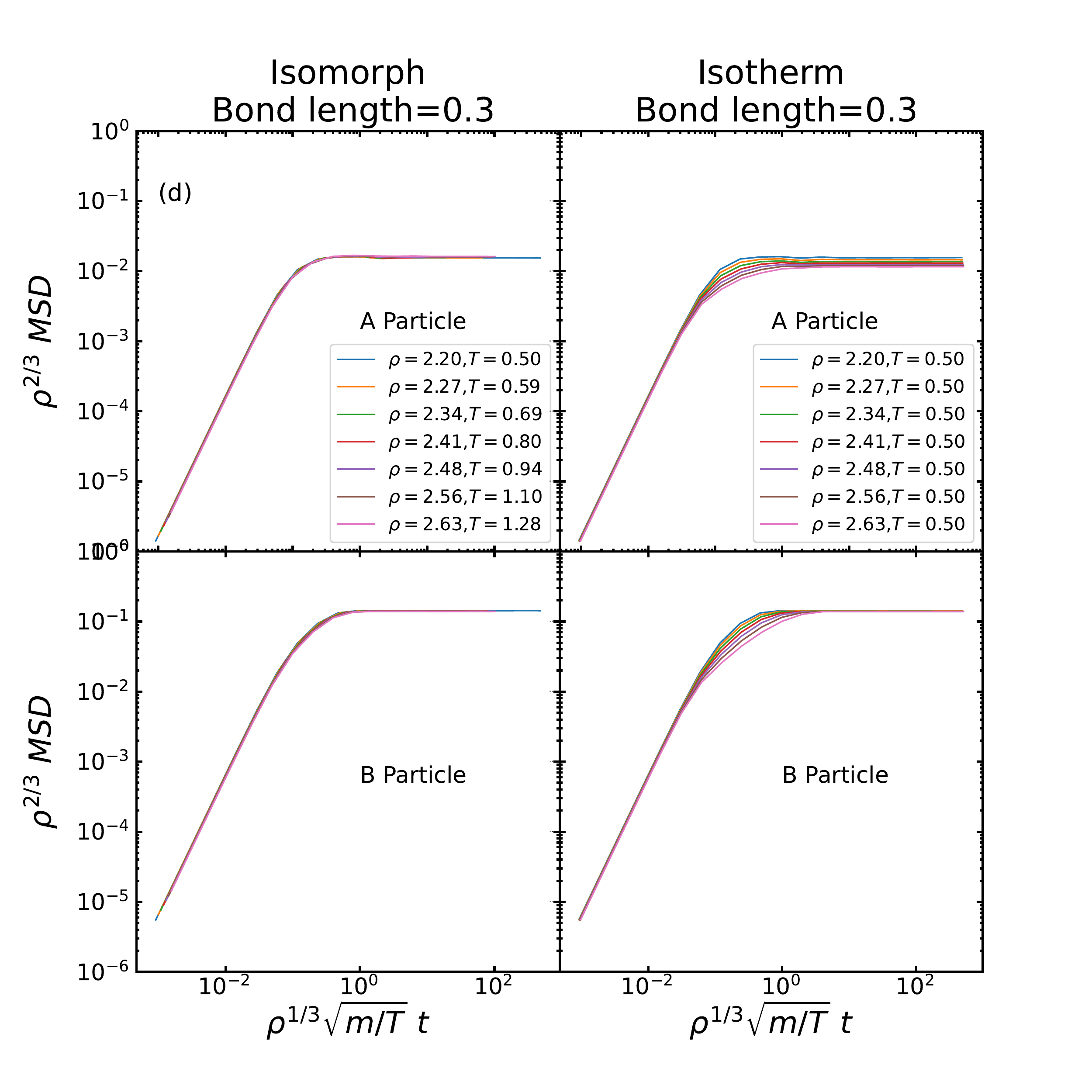} 
	\caption{(\textbf{a}) The A and B reduced-unit MSD along the isomorph and the isotherm for bond length $0.05$. There is a good invariance along the isomorph for both the A and B reduced MSD, but not along the corresponding isotherm. The short-time invariance in both cases derives from the definition of reduced units; the long-time plateau is far from invariant along the isotherm.
	(\textbf{b}--\textbf{d}) show similar plots for bond lengths 0.1, 0.2, and 0.3, respectively. Again there is invariance along the isomorphs for both MSDs, but not along the corresponding isotherms.}
  	\label{fig12}
\end{figure}

For the MSD as a function of reduced time (Figure~\ref{fig12}), we note that these are all isomorph invariant to a good approximation. In contrast, there is a notable variation along the reference-state-point isotherm for both A and B particles. Only at short times (in the ballistic regime) do we observe invariance along the isotherms, but as mentioned this is a consequence of the definition of reduced units.

\begin{figure}[H]
    \centering \includegraphics[width=7.5cm]{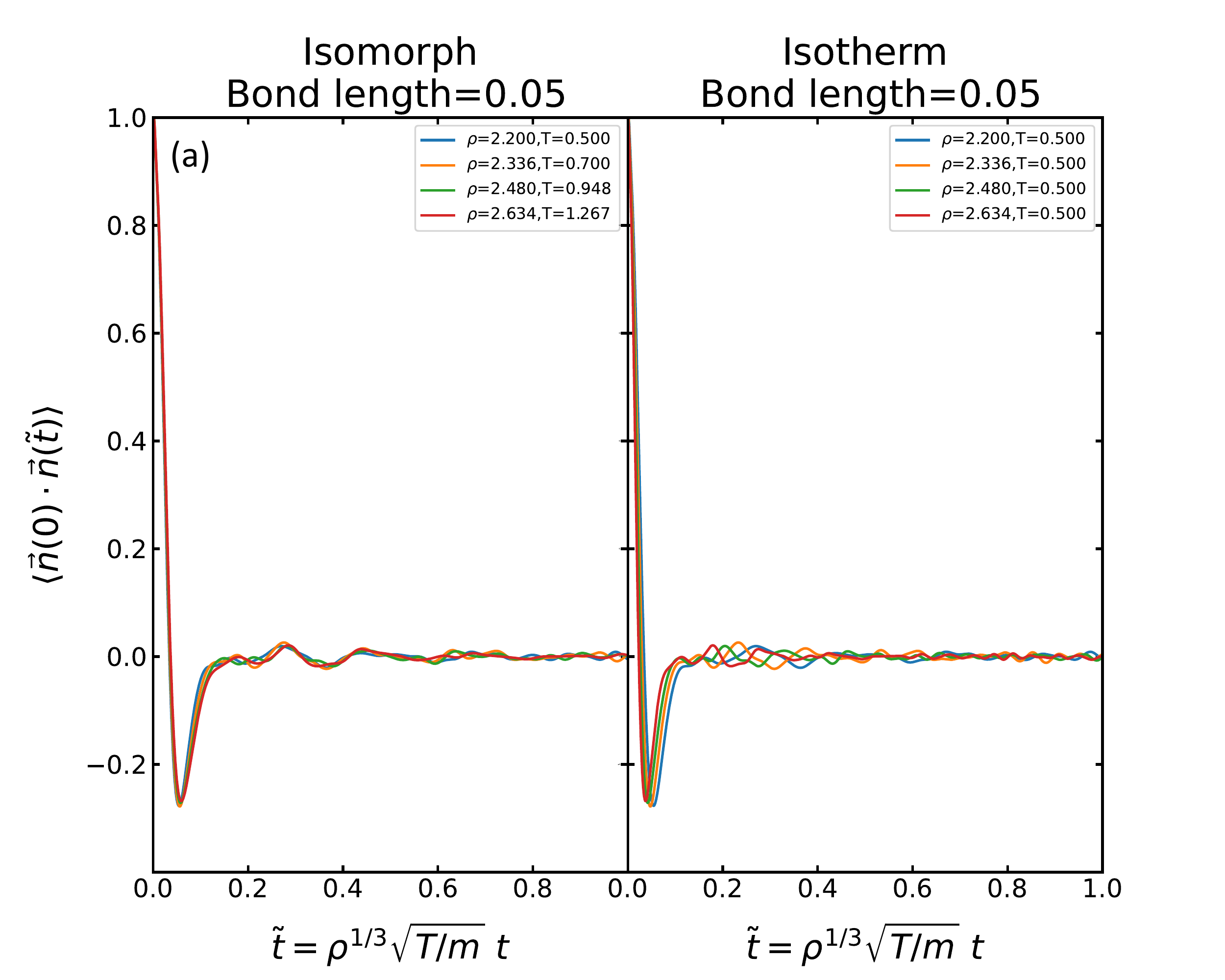}
  \includegraphics[width=7.5cm]{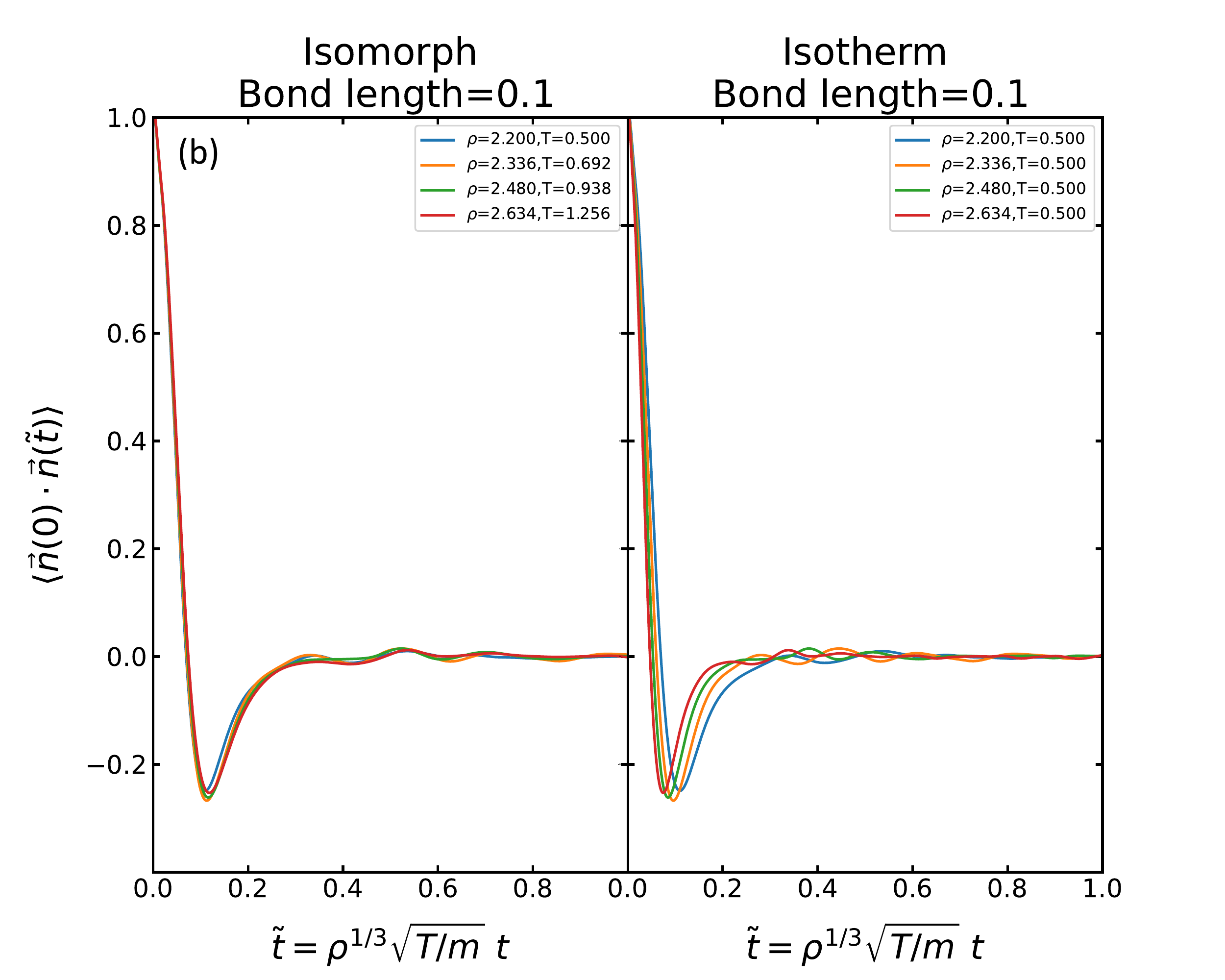}
  \includegraphics[width=7.5cm]{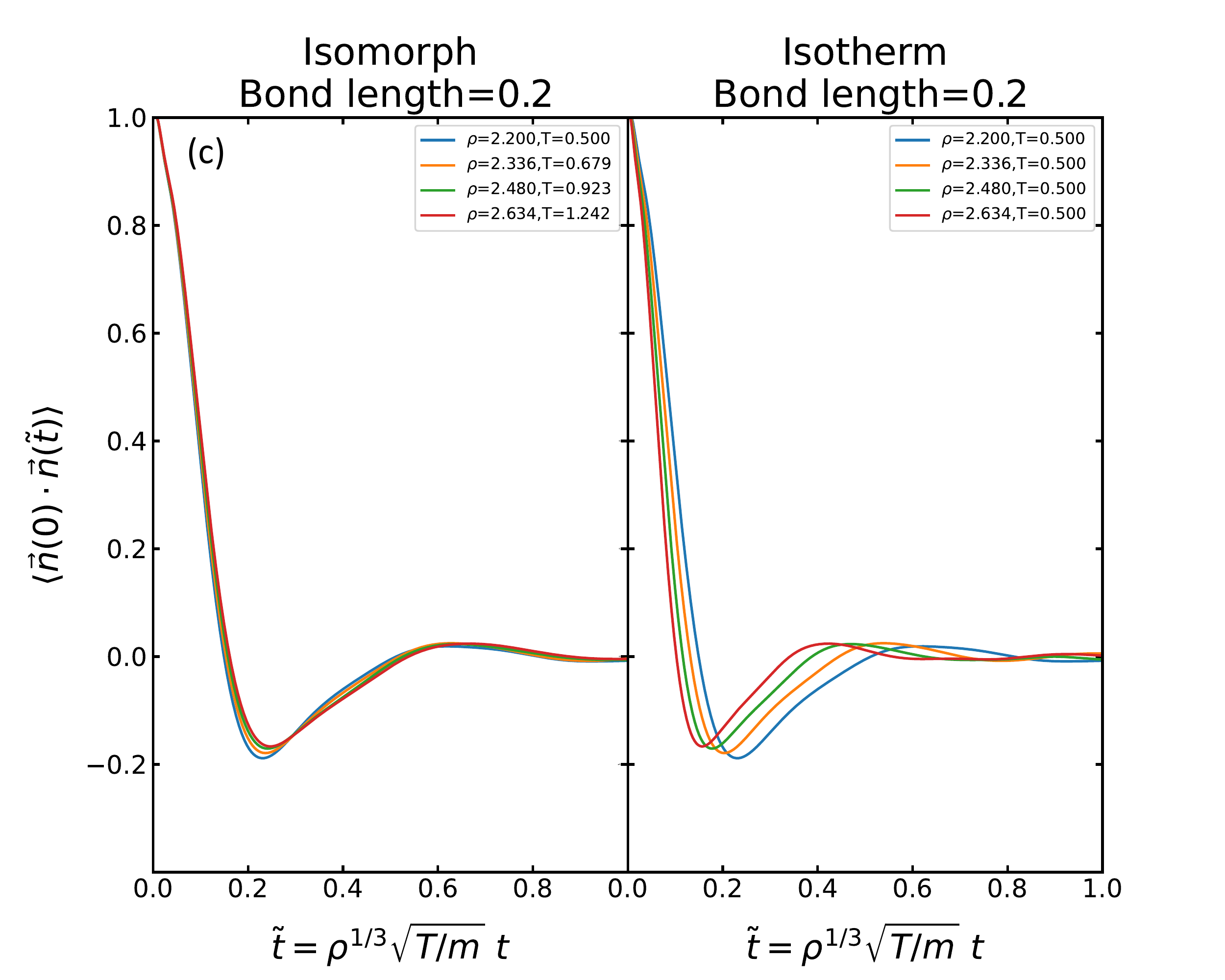}
  \includegraphics[width=7.5cm]{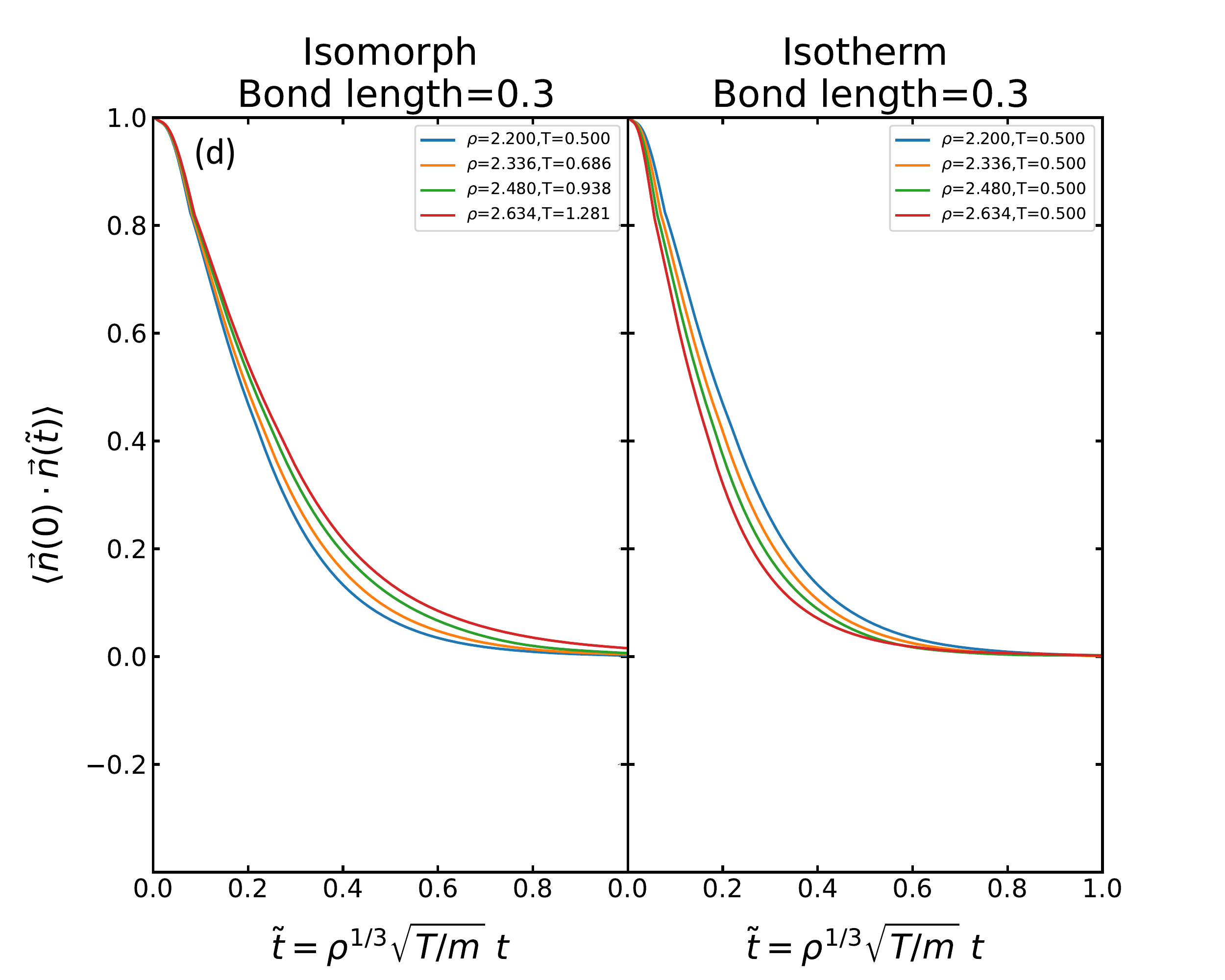}\vspace{6pt}
  \caption{(\textbf{a}--\textbf{d})
  	{The rotational}  time-autocorrelation function (RAC) along the isotherm  and the isomorphs for the four different bond lengths. As the bond length increases, the decay to zero becomes significantly slower. The largest bond length (0.3) behaves differently from the others by not going below zero. }
 \label{fig13}
\end{figure}

Proceeding finally to the RACs, Figure~\ref{fig13} shows these where the upper two presents the RACs along the isotherms for the four bond lengths and the lower row gives corresponding data for the RACs along the isomorphs. With the notable exception of bond length 0.3, we see a fairly good isomorph invariance of the RACs. Despite the already-mentioned fact that the RAC is not expected to be isomorph invariant because the reduced moment of inertia is not, we find for all bond lengths that the invariance along isomorphs is better than along the isotherms -- and much better in the bond-length 0.3 case.

\section{Summary}

We have presented a numerical study of the liquid and plastic-crystal phases of asymmetric dumbbell models of different bond lengths with the purpose of testing the isomorph-theory predictions of invariant structure and dynamics. In the liquid case, the bond lengths 0.05, 0.1, 0.2, and 0.5 were studied, while in the crystalline case the largest bond length was 0.3. At all state points the virial potential-energy correlation coefficients were above 0.84; in the vast majority of cases it was above 0.9, and for the plastic crystals it was quite close to unity. This implies that the isomorph theory is expected to apply~\cite{IV,dyr14}. Indeed, we found good isomorph invariance of the reduced-unit RDF and MSD in the liquid phase and even better invariance in the plastic-crystal phase. Isomorph invariance also applies to a good approximation for the RAC, despite the fact that this quantity is not predicted to be an exact invariant because the bond length is constant in real units, not in reduced units. In contrast, these three quantities were generally not invariant along isotherms of the same (20\%) density variation. The behavior of the B particles conform to a picture in which these are ``slaves'' of the A particles, which behave almost like single-component LJ particles, i.e., are not much affected by the B particles.

This paper is the first time the isomorph theory has been shown to apply for plastic crystals of simple two-atom dumbbell molecules. In conjunction with recent papers demonstrating the applicability of isomorph theory to liquid crystals~\cite{meh22,meh22a}, the present findings demonstrate the generality of this theoretical framework derived from the hidden-scale-invariance condition Equation~(\ref{hsi}) and, in particular, that the isomorph theory applies independent of which phase the system is in. It would be interesting to investigate the transition from the plastic crystal to the ordinary, orientationally ordered crystal phase and to study whether the phase transition line is close to an isomorph, as has been found for the liquid--solid transition line~\cite{ped16}.

\acknowledgments{This research was funded by the VILLUM Foundation's grant number 16515 for the \textit{Matter} project.}

\end{document}